# Evidence of Majorana fermions in an Al – InAs nanowire topological superconductor

Anindya Das[*], Yuval Ronen[*], Yonatan Most, Yuval Oreg, Moty Heiblum[#],
and Hadas Shtrikman

*Braun Center for Submicron Research, Department of Condensed Matter Physics,
Weizmann Institute of Science, Rehovot 76100, Israel*

[*]*equal contributions*
[#]*Corresponding Author (moty.heiblum@weizmann.ac.il)*

**Majorana fermions are the only fermionic particles that are expected to be their own antiparticles. While elementary particles of the Majorana type were not identified yet, quasi-particles with Majorana like properties, born from interacting electrons in the solid, were predicted to exist. Here, we present thorough experimental studies, backed by numerical simulations, of a system composed of an aluminum superconductor in proximity to an indium arsenide nanowire, with the latter possessing strong spin-orbit coupling. An induced 1d topological superconductor – supporting Majorana fermions at both ends – is expected to form. We concentrate on the characteristics of a distinct zero bias conductance peak (ZBP), and its splitting in energy, both appearing only with a small magnetic field applied along the wire. The ZBP was found to be robustly tied to the Fermi energy over a wide range of system parameters. While not providing a definite proof of a Majorana state, the presented data and the simulations support strongly its existence.**

**Background**

Quantum mechanics and special relativity were merged into a single theory when Dirac presented his equation in 1929 [(1)], with a solution predicting an electron and an anti-electron partner - the positron. All fermionic particles have their own anti-partners. Majorana, however, showed that Dirac's equation has also real solutions - the so-called Majorana fermions [(2)]. Majorana fermions are unique because they are their own anti-particles. While in high-energy physics the Majorana is an elementary particle [3], in condensed matter physics one of its manifestations is an emergent quasi-particle zero energy state. In 1d, two of such states are expected to appear at the edges of a 'spinless' (namely, with a single band) superconductor. We refer to these states for short as 'Majoranas' [4], [5].

The fundamental aspects of Majoranas, and their non-abelian braiding properties [6], [7], [8], leading to possible applications in quantum computation [9], [10], [11], made the search for them a recent popular field of study in condensed matter physics. Examples of leading candidates, which under the right conditions may host these quasi-particles, are: (*i*) Read-Green type states of the fractional quantum Hall effect (FQHE) [12]; (*ii*) Vortices in 2D *p+ip* spinless superconductors [13]; and (*iii*) Domain walls in 1d P-wave superconductors [14], [15]. Since utilizing conventional S-wave superconductors is quite appealing, as they are more easily implemented than P-wave superconductors, several suggestions for such implementations were recently proposed: (*i*) Surface of a 3d topological insulator in proximity to an S-wave superconductor [16]; (*ii*) 2d semiconductor with strong spin-orbit coupling in proximity to an S-wave superconductor under broken time reversal symmetry (say, via a local ferromagnet [17] or magnetic field [18]); and, (*iii*) 1d semiconductor (similar configuration to the above) with the Majorana quasiparticles appearing at the two ends of the 1d wire [19], [20]. Specifically, the latter authors proposed to employ InAs or InSb nanowires, possessing strong spin-orbit coupling as well as large Zeeman splitting at low magnetic fields, in proximity to an S-wave superconductor. Following the suggestion of Oreg *et al*. [20] to use an InSb nanowire, Mourik *et al*. [21] and Ref [22] have recently reported the observation of a magnetic field dependent zero-bias peak in the conductance; as expected for a zero energy Majorana fermion state.

In this paper, we report the observation of a zero bias-peak (ZBP) and its splitting, at a finite magnetic field, in a high quality InAs nanowire placed in proximity to a superconducting Al electrode. We study the dependence of the peak height and width on the system's parameters, such as, chemical potential, potential barriers, magnetic field and its orientation, and the electrons' temperature. We compare the experimental results with numerical simulations based on scattering theory and find, using the experimental parameters, a qualitative agreement of the data with a Majorana state. We also discuss alternative models that may account for the observed ZBP [23], [24], [25], [26].

**Theoretical aspects**

It is important to specify the required conditions for the formation of a Majorana state in a solid. The basic requirements of a quasi-particle being identical to its anti-quasi-particle is that it is spinless (single pseudo spin) and has zero charge. These requirements can be satisfied by unique Cooper pairing in a P-wave superconductor [5], or in filling factor $n+\frac{1}{2}$ (with $n$ integer) of the FQHE, where two same-spin composite fermions condense into a Cooper pair-like quasi-particle [12]. We present a realization of a 1d nanowire coupled to an S-wave superconductor, hence, with induced superconductivity in the wire (a manifestation of the proximity effect [27]). Moreover, a Rashba spin-orbit coupling [28], which leads to an effective magnetic field $\vec{B}_{so} \propto \vec{p} \times \vec{E}$, separates electrons having opposite spins in momentum space (momentum $\vec{p}$ is along the wire; gate induced electric field $\vec{E}$ is perpendicular to the wire). Without superconductivity, applying an external magnetic field perpendicular to $B_{so}$ will mix the two spin bands, forming two 'pseudo spin bands' separated by an energy gap (Zeeman gap, $E_Z$) at $p=0$ (Figs. 1a & 1b, see also Figs. 7a&7b). Inducing superconductivity modifies the $p=0$ gap and opens up another gap at the Fermi momentum $p_F$ (Fig. 1c). The overall gap $E_g$ is the smaller of these two gaps. Three parameters are of significance: the spin-orbit energy, $\Delta_{so} = p_{so}^2/2m$, with $\pm p_{so} = \pm \hbar/\lambda_{so}$ the corresponding momentum to the bottom of the spin-split band; the Zeeman gap $E_z = \frac{1}{2} g\mu_B B$, where $g$ the Lande $g$-factor, $\mu_B$ the Bohr magnetron, and $B$ the external magnetic field; and the induced superconducting gap in the nanowire $2\Delta_{ind}$ (smaller than the Al superconducting gap, $2\Delta_{Al}$). For $\Delta_{ind}>0$ and $E_z = 0$ the wire is in the trivial superconductor phase and its spectrum is gapped. As we increase $E_z$ the gap closes at one isolated value of the Zeeman energy, $E_z = \sqrt{\Delta_{ind}^2 + \mu^2}$, with µ the chemical potential (µ=0 is at the two spin bands crossing). At this value of $E_Z$, the gap closes at the Fermi energy at $p=0$ (and a 'topological phase transition' occurs). At higher values of $E_z$ the wire is gapped again – this is the topological superconductor phase. When we continuously change the phase of the wire from any other gapped phase into the topological phase, we need to close the gap somewhere along the way. Hence, when a finite segment of the wire is in the topological phase, the gap must close at its ends, forming two Majorana states – each at one end of the segment [29]. The two Majoranas can combine into a single nonlocal fermion.

The required condition for a topological superconductor is therefore $E_z > \sqrt{\Delta_{ind}^2 + \mu^2}$ (or equivalently $E_z > \Delta_{ind}$ and $|\mu| < \sqrt{E_z^2 - \Delta_{ind}^2}$). For Zeeman energies close to the topological phase transition, the minimal gap is at $p=0$: $E_g = 2\left|E_z - \sqrt{\Delta_{ind}^2 + \mu^2}\right|$ (Fig. 1f). For larger Zeeman energies, the gap at $p_F$ is the smaller of the two, which for $E_z \gg \Delta_{so}$, $\Delta_{ind}$, $|\mu|$ is given by $E_g = 4\Delta_{ind}\sqrt{\Delta_{so}/E_z}$ (Fig. 1f). In this regime the gap decreases with $E_z$, and even more so due to the diminishing of the superconducting gap with field. The Majoranas, with energy $\varepsilon$ pinned to the Fermi energy, should be robust for a range of system parameters that keep $|\mu| < \sqrt{E_z^2 - \Delta_{ind}^2}$. The immediate consequence of the Majorana is a sharp enhancement in the tunneling density of states and a $2e^2/h$ differential conductance at zero applied bias and $T=0$ [30], [31] (the actual width and height of the peak are determined by the relation of the barrier height to the electron temperature, and the coupling between the two Majoranas at the ends of the topological segment).

**Setup of devices**

An SEM image of our device, as well as its schematic illustration, is shown in Fig. 2. Stacking faults free, 50-60nm diameter Wurtzite InAs nanowires, were grown by Au-assisted vapor-liquid-solid (VLS) MBE process on (011) InAs substrate after oxide blow off and in-situ evaporation of a thin gold layer in a vacuum connected chamber [32]. A typical high resolution TEM image taken from such nanowire is shown in the inset of Fig. 2 (see Fig. S1). The nanowires were suspended (some 50nm high) above an oxidized conducting Si substrate with a 150nm thick $SiO_2$ layer. Two types of devices were tested (Fig. 2a & 2b); in both the wire was suspended on two gold pillars at both ends with gold layers on top serving as low resistance contacts. A superconducting aluminum thin strip, 100nm thick and ~150nm wide, was intimately contacted at the center of the nanowire, serving as source, separating the wire into two equal sections, each ~200nm long. In type I device a gold pillar supported the wire under the aluminum electrode (critical field of Al~60-70mT), while in type II device the center pillar was missing (critical field of Al varies from ~ 100-150mT for different devices). The conducting Si substrate served as a global gate (GG) for the entire wire (only in type II), while two additional narrow local gates (LG & RG, Fig. 2d), placed 80nm away from the superconductor edge (25nm thick, 50nm wide), strongly influenced the barrier height as well as the chemical potential in the wire.

Before cooling, the devices dwelled at room temperature in a vacuum pumped chamber for some 24 hours in order to remove surface impurities, with the conductance increasing by some twenty-fold. With the dilution refrigerator temperature at 10mK, the estimated electron temperature in the wire was ~30mK (see later). A 575Hz 1-2µV RMS signal was fed to the superconducting contact and the resultant current was collected at one side of the wire by the drain contact, later to be amplified by a homemade current amplifier (see Fig. 2). We also measured the conductance at a higher frequency (~1MHz), employing a low noise voltage cold preamplifier cooled to 1K (see also Ref [33]).

Our numerical simulations use a generalization of the formalism pioneered by Blonder, Tinkham and Klapwijk (BTK) [34], which allows modeling a large number of segments of the wire, includes spin flip processes, and goes beyond the small bias approximation. Each segment is characterized by different parameters, which jump discontinuously at the interfaces between the segments. Using wavefunction matching at the interfaces, we write the Bogoliubov-de Gennes (BdG) equations [27], and solve them to find the exact scattering states at each energy, and thus the corresponding transmission and reflection amplitudes (for further details see the Supplementary Section). The similar approach has been used in the Refs [35], [36], [37], [38].

**Study of parameters**

We start with a 'calibration' of the two types of the studied devices. Bare and un-gated wires are n-type with density $\sim 10^6 \text{cm}^{-1}$, and thus are likely to occupy a single sub-band. The presence of disorder makes the conductance highly sensitive to electron density, namely, gates voltage. At the lower conductance range, Cooper pair transport is suppressed and the zero bias conductance exhibits either dips ($\sim 0.5 e^2/h$) or ZBPs. The observed peaks are likely to result due to the following effects: (*i*) 'Reflectionless scattering', being constructive interference between electron reflection and Andreev reflection [39] in S-I-N-I devices (where I - insulator, S - superconductor, N - normal)- which is expected to be suppressed with *B* field [40], [41], [42]; (*ii*) An Andreev bound state in a S-N-I, which is likely to be split at zero field - however such a state is expected to exhibit Zeeman splitting [43], [44]; (*iii*) Kondo correlations in a strongly coupled confined electron puddle between the wire and the leads (I-N-I) – which is expected to exhibit Zeeman splitting [45], [46]; and (*iv*) Kondo-like peak appearing when superconductivity is quenched. From the Zeeman split ZBP, we estimate that *g*~20 (we assumed an equal voltage drop on the two potential barriers at the superconductor-normal and ohmic contact-normal interfaces, and thus $4E_z \cong 125 \mu eV$ at *B*=50mT) [47]. The apparently large *g*-factor may also indicate an enhanced *B* field in the wire due to its repulsion from within the superconductor. These multiple ZBPs or their absence are discussed in some detail in the Supplementary Section.

Most of the presented data was taken with type II devices, where the main features in the non-linear conductance (see Figs. 3-5): two symmetric weak shoulders, being the 'induced gap' in the nanowire under the superconductor $\Delta_{ind} \cong 50 \mu eV$; and distinct peaks, being the aluminum gap $\Delta_{Al} \cong 150 \mu V$. We also made a few measurements on type I devices, where shot noise measurement (at ~1MHz) revealed an abrupt change in the slope at $eV_{SD} = \Delta_{Al}$. For a transmission coefficient *t*~0.6, current *I*, and *T*~10mK, the measured noise spectral density agreed with $S \sim 2e^* I(1-t^*)$, with $e^* = 2e$ and $t^* = t^2$ for $eV_{SD} < \Delta$ and $e^* = e$ with $t^* = t$ and $eV_{SD} > \Delta_{Al}$ [33], [48]. Failing to observe the induced gap in this measurement reaffirms that the dominant process is of pairs tunneling into the superconductor below $\Delta_{Al}$.

**Characterization of the field emerging zero bias conductance peaks**

The gold pedestal under the aluminum contact in type I device fully screened the global gate voltage, thus preventing tuning the chemical potential under the aluminum contact. Indeed, none of the features observed in type II had been observed in two separate type I devices (D1&D2).

Measurements were done on two separate type II devices (D3&D4), and one of them thermally cycled a few times. The differential conductance $G_R$ was measured between the superconducting contact and right normal contact, while keeping the left side of the wire pinched. For a fixed $V_{GG}$, the conductance increased rather abruptly with increasing $V_{RG}$. Coulomb diamonds indicated in the pinched-off region a distinct Coulomb blockade behavior, followed by a Kondo correlated regime as the voltage increased, and finally to Fabry-Perot fluctuations with an average conductance of $G\sim 2e^2/h$ [49].

In order to locate the ZBP at finite magnetic field, we measure the $G_R$ ($V_{SD}$=0) as a function of $V_{RG}$ for different negative values of $V_{GG}$ at zero and at ~50mT. In general, at 50mT the conductance gets smaller, however, at sufficiently large negative $V_{GG}$ (~ -15V to -18.5V, among different devices) the zero bias conductance was observed to increase at 50mT – identifying the presence of a ZBP (see a demonstration in Supplementary Section). We discuss now in some detail the data measured on device D4, which is in the heart of the paper. (ZBP appearing at (Fig. 3a)),

We start with Fig. 3a, with $V_{GG}$=-18.3V and $V_{RG}$=1.17-1.24V, where $V_{RG}$ is controlling mainly the chemical potential under the superconductor. The conductance in the left panel, $B$=0, exhibited an obvious dip, flanked by two shoulders at $\pm V_{SD}$~45μV, which are interpreted as the induced gap $\Delta_{ind}$ - appearing to weakly change with $V_{RG}$. With $B$=30mT and at $V_{RG}$=1.205V (center broken line), the two conductance shoulders came closer and turned into a relatively wide ZBP, barely split (peak with a shoulder). At higher gate voltage, $V_{RG}$=1.183V, the split peak appears as two weak shoulder around $V_{SD}$=0. For $B$=50mT, the observed ZBP at mid gate voltage range is seen to be more distinct peak (height $G_{peak} \cong 0.1e^2/h$ and width ~18μV - approximately agreeing with an electron temperature of 30mK); persisting for a wider range of gate voltage. Splitting is clear at the high and low gate voltage range. In the right panel, $B$=70mT (approaching the Al critical field, 100mT), the conductance features were weaker, with the split peaks in most of the gate voltage range. See the Supplementary Section for more B fields. The results can concisely be summarized in Fig. 3b, where we plot the conductance of the ZBP (after background subtraction) as a function of $V_{RG}$ and $B$. Each contour line indicates an equal height of ZBP, which can be also approximated (assuming weak effect on the potential barrier) as an equal wavefunction extent of the Majorana. The three arrows indicate the region of transition between single ZBP and split peaks.

We dwell on this point some more now. We have shown in the Theoretical Aspects Section that the wire enters the topological phase for $E_z > \Delta_{ind}$, but only within $\Delta\mu = 2\sqrt{E_z^2 - \Delta_{ind}^2}$. Namely, the wire is in the topological phase only for a limited range of the chemical potential $\mu$, and when $\mu$ is close to the edge of that range, the gap is small and the overlap of the Majoranas causes the split (Majorana wave function spatial length is $\xi \approx \frac{\hbar v_F}{E_g}$, with $v_F$ being the Fermi velocity). Hence, the presence of a single peak in the $\Delta\mu(B)$ suggests a relatively large gap $E_g$ and a short extent of the Majorana wave function – preventing splitting of the ZBP. This contour is expected to follow the expression for $\Delta\mu$ at the low range of $B$, given above for $E_g(0)$. Alternatively, at high $B$, the dominant energy gap $E_g(p_F)$ is at the Fermi momentum, and thus its extent is barely affected by μ. Hence, the ZBP should remain unsplit as long as the superconductor gap is large enough. This is seen in Fig. 3b, where the contour climb resembling square root dependent with 'sudden death' around 70mT, with the ZBP is always split. We performed a theoretical analysis of the estimated extent of the Majoranas wave function on μ and $B$ and compared it with the length of the topological superconductor ($L$=150nm). Figure 3c presents three contours for different idealized wave functions extent, where a single ZBP is expected for $\xi < 3L$ (red line in Fig. 3c, note that $\xi$ was calculated for semi-infinite wire). Note, that we presented the above comparison in order to highlight the resemblance of the functional dependence of the contours. While the robustness of the ZBP with $V_{RG}$ is clearly demonstrated in Fig. 3, one can also test its rigidity as a function of $V_{GG}$, which is shown in the Supplementary Section.

We continued to explore, in some more detail, the dependence of the ZBP and its splitting as function of magnetic field in three extreme regions of the chemical potential. We started with the highest gap, namely, $\mu \sim 0$ ($V_{RG}=1.205$V in Fig. 3a). The magnetic field was scanned from $B=0$ upwards (Figs. 4a & 4b). The two inner shoulders (being $\Delta_{ind} \sim 45\mu$eV at $B=0$) move closer in energy with $B$, and merge into a ZBP at $B \sim 35$mT. The ZBP splits beyond $B \sim 70$mT (Figs. 4a - 4c); however, to a less extent. It is important to note that the low energy features disappeared with the superconducting gap collapsing at $B \sim 100$mT. See more data $V_{RG}=1.21$V in the Supplementary Section.

Is this behavior understood in the light of Majoranas quasi-particles? The approximate appearance of a single ZBP at $B \sim 35$mT is somewhat higher than that expected to close the gap (with $g=20$ we expect $B \sim 30$mT). This suggests that features of a closing gap $E_g(0)$ merger with a split ZBP in the range $B=30-35$mT. At the high field region, the splitting of the ZBP results from shrinking gap at the Fermi momentum $E_g(p_F)$. We performed numerical simulations for a single sub-band wire, where many of the system parameters were taken into account, except for disorder (details in Supplementary Section). Assuming spin orbit energy $\Delta_{so} \sim 70\mu$eV (reported values are $\lambda_{so} \approx 100nm$ giving $\Delta_{so} \approx 50\mu$eV [50], [51]) and $v_F \propto \sqrt{\Delta_{so}}$ (leading to $\xi \approx \frac{\hbar v_F}{E_g} \sim 285$nm), with an effective distance between the two Majoranas ~160nm (very close to the superconductor width, $L$). The simulations results are shown the color scale conductance in Fig. 4d – with a rather good agreement with the data.

We moved the chemical potential to its higher end, $V_{RG}=1.224$V (Fig. 3a); expecting to further reduce the topological gap and enhance splitting of the ZBP. The color scale and the corresponding cuts in Fig. 5 show a split peak in all regions of the magnetic field; the split peaks remain parallel for a relatively wide extent of $B$. The split peaks die when the superconducting gap vanishes. The representation of the chemical potential in the lower range is given in the Supplementary Section.

The validity of assigning the ZBP to the Majorana quasi-particle can be further tested by measuring the sensitivity of the peak to the temperature and to the orientation of the magnetic field. Temperature dependence of the type II device (D3, with a critical field >150mT) at $B$=70mT is shown in Fig. 6. Fig. 6a provides constant temperature cuts of the conductance as a function of $V_{SD}$. In Fig. 6b, the peak height $\Delta G_{max}$ and width (~18μV at the lowest temperature) are plotted as a function of the lattice temperature, with the peak disappearing at $T_l$~100mK. The actual peak height may deviate from the expected value of $2e^2/h$ due to: Coupling of the Majorana fermions due to Coulomb interactions and tunnelling [52], [53], masking by the background non-vanishing conductance, or due to temperature averaging. Assuming the latter, the width seems to provide a reasonable measure of the electron temperature $T_e$. At base temperature of the lattice ~10mK, we found $3.5k_BT \cong 9\mu eV$ [54], suggesting an electron temperature of $T_e$~30mK, which is quite reasonable. Initially, the peak height does not follow $G_{max} \propto \Gamma/T_e$ ($\Gamma$ - natural energy broadening of the ZBP), since the change in the electron temperature lags behind that of the lattice, however, a crude estimate leads to $\Gamma$~1μeV.

Since the topological gap opens up when the applied external field is perpendicular to the spin orbit field, we expect the ZBP to disappear when the Zeeman field is parallel to the spin-orbit effective field. To test this dependence, the sample (D3) was placed on a piezoelectric rotator (Autocube # ANR220\RES) and the evolution of the ZBP was plotted in Fig. 7, for a few angles in the wire – spin orbit direction plane. The ZBP nearly disappeared at some 75° (0° is along the nanowire).

**Discussion**

Our study of the composite InAs nanowire - aluminum superconductor system, supports the formation of a topological superconductor with Majorana states at its ends. We summarize the main features. The ZBP… (*i*) was *not* found in type I devices (D1&D2), where the chemical potential could not be tuned (screening due to a metal pedestal); (*ii*) appeared only in a limited range of the chemical potential, and was robustly stuck to the Fermi energy; (*iii*) appeared only right after the Zeeman gap exceeded somewhat the induced gap in the wire; (i*v*) was observed to split when the gap in the topological phase was small (hence, the extent of the wave-function large); (*v*) decreases with temperature and finally disappeared at ~100mK; (*vi*) decreased and disappeared with rotation of the magnetic field towards the spin-orbit direction.

What can we say about other possible scenarios that may also lead to some of the observed behavior? (*i*) The weak anti-localization scenario [55] can be ruled out as the normal segment of the wire had a single channel and the ZBP strongly dependent on the orientation of magnetic field. (*ii*) We believe that conventional Kondo effect can be ruled out for a variety of reasons: (a) The normal part of the wire is tuned beyond the quantum dot regime (Supplementary Section); (b) The capacitance of the topological wire (due to the superconductor coating) is very large, leading to a small charging energy (in any likely quantum dot in the topological segment); (c) Even for a *g*-factor ~2, the splitting at ~100mT would be ~25µeV, and thus should be observable; (d) The ZBP disappears as the superconductor gap quenches with magnetic field. However, our experimental findings cannot rule out exotic scenarios such as orbital Kondo effects. (*iii*) Finite energy Andreev bound states may move and anti-cross each other [56], however, they may be excluded using the same argument of the sizable *g*-factor in our device. (*iv*) The current observations cannot rule out the possibility of clusters of low-energy states being localized near the wire end due to disorder in a multichannel wire below the superconductor [24]. Nevertheless, we would like to point out that we believe that the elastic mean free path in the wire was found to be longer than the length of the wire [49], and no evidence of peaks clustering was found.

To further study the properties of these systems additional experimental and theoretical investigations should be conducted. We would like to point out that the non-local Josephson effect [57] and shot noise experiments [58] may help to distinguish between the different scenarios.

## Acknowledgements

We wish to thank A. Haim, A. Stern, F. von Oppen and G. Refael for useful discussions. We are grateful to R. Popovitz-Biro and D. Mahalu for professional contribution and to A. Kretinin for laying the ground for doing nanowires device physics, and S. Ilani and A. Joshua for enabling us to perform the tilted field measurement. MH acknowledges the partial support of the Israeli Science Foundation (ISF), the Minerva foundation, and the US-Israel Bi-National Science Foundation (BSF), the European Research Council under the European Community's Seventh Framework Program (FP7/2007-2013) / ERC Grant agreement # 227716. YO acknowledges the partial support of DFG, Minerva and that of the BSF. HS acknowledges the partial support of ISF and Israeli Ministry of Science and Technology (IMOST).

**Figure Captions**

**Figure 1.** Energy dispersion of the InAs nanowire excitations (Bogoliubov-de Gennes spectrum), in proximity to the Al superconductor. Heavy lines show electron-like bands and light lines show hole-like bands. Opposite spin directions are denoted in blue and magenta (red and cyan) for the spin-orbit effective field direction (perpendicular direction), where a relative mixture denotes intermediate spin directions. (a) Split electronic spin bands due to spin-orbit coupling in the InAs wire. Spin-orbit energy defined as $\Delta_{so}$, with the chemical potential $\mu$ measured with respect to the spin bands crossing at $p=0$. (b) With the application of magnetic field perpendicular to the spin-orbit effective magnetic field ($B_{so} \perp B$), a Zeeman gap, $E_z = \frac{1}{2} g\mu_B B$, opens at $p=0$. (c) Light curves for the hole excitations are added, and bringing into close proximity a superconductor opens up superconducting gaps at the crossing of particle and hole curves. The overall gap is determined by the minimum between the gap at $p=0$ and the gap at $p_F$, while for $\mu=0$ and $E_z$ close to $\Delta_{ind}$ the gap at $p=0$ is dominant. (d) As in (c) but $E_z$ is increased so that the gap at $p_F$ is dominant. (e) $B$ is rotated to a direction of 30° with respect to $B_{so}$. The original spin-orbit bands are shifted in opposite vertical directions, and the $B$ component, which is perpendicular to $B_{so}$ is diminished. (f) The evolution of the energy gap at $p=0$ (dotted blue), at $p_F$ (dotted yellow), and the overall energy gap (dashed black) with Zeeman energy, $E_z$, for $\mu=0$. The overall gap is determined by the minimum of the other two, where the $p=0$ gap is dominant around the phase transition, which occurs at $E_z = \Delta_{ind}$. At high $E_z$ the $p_F$ gap, which is decreasing with $E_z$, becomes dominant.

**Figure 2.** Structure of the Al-InAs structures suspended above p-type silicon covered with 150nm $SiO_2$. (a) Type I device, the nanowire is supported by three gold pedestals, with a gold 'normal' contact at one edge and an aluminum superconducting contact at the center. The conductive Si substrate serves as a global gate (GG), controlling barrier as well as the chemical potential of the wire. Two narrow local gates (RG and LG), 50nm wide and 25nm thick, displaced from the superconducting contact by 80nm, also strongly

influence the barrier height as well as the chemical potential in the wire. (b) Type II device, similar to type I device, but without the pedestal under the Al superconducting contact. This structure allows controlling the chemical potential under the Al contact. (c) SEM micrograph of type II device. A voltage source, with 5 Ohm resistance, provides $V_{SD}$, and closes the circuit through the 'cold ground' (cold finger) in the dilution refrigerator. Gates are tuned by $V_{GG}$ and $V_{RG}$ to the desired conditions. Inset: High resolution TEM image (viewed from the <1120> zone axis) of a stacking faults free, wurtzite structure, InAs nanowire, grown on (011) InAs in the <111> direction. TEM image is courtesy of Ronit Popovitz-Biro. A more detailed image can be found the Supplementary Section. (d) An estimated potential profile along the wire. The two local gates (LG and RG) and global gate (GG) determine the shape of the potential barriers; probably affect the distance between the Majoranas.

**Figure 3.** Dependence of the ZBP on the chemical potential and the magnetic field. (a) Evolution of the ZBP for different magnetic fields as function of $V_{RG}$ (range from 1.17 to 1.24V) and bias for type II device (D4) at $V_{GG}$=-18.3V. $V_{RG}$ affects strongly the chemical potential under the superconductor. The cuts are taken at three local gate voltages (1.83V, 1.205V and 1.228V). At $B$=0mT, the bulk superconducting Al gap ($\Delta_{Al}$~±150μeV) with two inner shoulders at the induced gap, $\Delta_{ind}$~±45μeV. For $B$=30mT, the two conductance shoulders merge at $V_{RG}$=1.205V, and turn into a relatively wide, barely split, ZBP, which splits at higher and lower gate voltages. For $B$=50mT, the ZBP is a sharper single peak; persisting for a wider range of gate voltage, however, splitting is evident at the lowest and highest $V_{RG}$. This is likely to be a direct result of the increasing chemical potential, which lowers the gap, $E_g(0)$, and increase the coherence length of the Majorana, thus enhancing coupling and splitting of the ZBP. For $B$=70mT, the ZBP peak is split for a wide range of gate voltage. This is a consequence of overlapping between two Majorana. At higher magnetic field the gap is determined by $E_g(P_F)$, which decreases with $B$. (b) 2D color plot of the appearance of the ZBP, with contours lines of the equal height of ZBP, from $0.106e^2/h$ to $0.197e^2/h$. Three arrows show the transition from split peaks to single ZBP. (c) Using analytical expressions for the wire spectrum, contours lines of constant estimated size of the wave function of the Majorana, $\xi =$

$\hbar v_F/E_g$, (~1.5$L$, 3$L$ and 10$L$ for blue, red and black line, respectively) plotted as a function of the magnetic field $B$ and the chemical potential $\mu$. Only the region where the wire is in the topological phase (and supports Majoranas at its ends) is plotted. We expect the ZBP to be observed when $\xi < \alpha L$, where $L$ is the length of the topological segment and $\alpha$ is some unknown constant due to the details of the Majorana wavefunction and finite size as well as finite temperature effects, which may modify the spectrum. Indeed the shape of the $\xi < 3L$ (red line) contour is similar to the contours shown in (b). While the range of $\mu$ for which the wire is topological increases as a function of $B$, for higher values of $B$ the gap decreases (due to the decreasing of $\Delta_{ind}$). This makes the Majoranas spatially wider and reduces the range of $\mu$ for which the ZBP is expected to be observed. The sharp termination of each contour at some maximal value of $B$ is explained by the fact that $\Delta_{ind}$ – depends weakly on $\mu$.

**Figure 4.** Low energy non-linear conductance as function of applied magnetic field, applied parallel to the wire axis (perpendicular to the effective spin-orbit magnetic field) of the device D4. (a) Color plot and (b) & (c) Cuts for every $\Delta B$~2mT, each shifted by~0.02$e^2/h$. The characteristic behavior, exhibits a relatively sharp conductance dip at $V_{SD}$=0 and $B$=0, flanked by two shoulders at $\pm eV_{SD}$~45μV, which are interpreted as the induced gap $\Delta_{ind}$. Outer peaks at $\pm V_{SD}=\Delta_{Al}\cong$150μV, are the aluminum gap $\Delta_{Al}$. With increasing $B$, the two shoulders come approach each other and merge to a single ZBP at around ~30mT and remain so until ~70mT. At ~50mT the ZBP is the sharpest, having a maximum value of ~0.12$e^2/h$. Beyond ~70mT the ZBP splits and the conductance features become weaker with increasing $B$. In (c) a zoom in of the cuts between ~65-120mT. The split peaks remain almost parallel with increasing $B$ and die before ~95mT; where the superconducting gap is still visible. At ~120mT conductance becomes featureless. (d) A simulation of the conductance with a topological segment length of 160nm and spin-orbit energy of $\Delta_{so}$=70μeV. The length was chosen to be comparable to the estimate of the separation between the two potential barriers (~150-200nm). A second channel was added in parallel, which ends with $\Delta_{Al}$ superconductor, to account for the quasiparticles tunneling into the aluminum at high energy (the contribution of the second channel was ~75% while that of the main channel was ~25%, see Supplementary Section). The measured dependence of $\Delta_{Al}$ and $\Delta_{ind}$ on the magnetic field was used. The data was convolved with a Fermi-Dirac kernel to simulate an electron temperature of 30mK.

**Figure 5.** Low energy non-linear conductance with a high chemical potential. (a) 2D color plot, and (b) Cuts with bias and magnetic field for $V_{RG}$=1.224V. The two shoulder peaks come closer with *B* but remain split, and parallel, in a wide region of the magnetic field (from ~60mT to 85mT). They disappear when the Al superconductivity is quenched. (c) A zoom in of the cuts in the interval ~65mT - 120mT, making observation of the splitting clearer. At ~120mT the conductance becomes featureless.

**Figure 6.** Temperature dependence at *B*=70mT of the ZBP of device D3. (a) Cuts for every $\Delta T$=8mK shifted by $0.05e^2/h$ each. The peak vanishes before *T*~100mK. From the summary of the data shown in (b), one can estimate (using the peak width) that the electron temperature is some ~30mK higher than that of the dilution refrigerator. Moreover, the temperature is responsible for the smallness of the peak as well as determining its effective measured width.

**Figure 7.** Changing the relative angle between the magnetic field and that of the spin-orbit effective field. (a) Band diagram for external field perpendicular to spin-orbit effective field, where one should expect to observe the maximum ZBP. (b) When an external field parallel to the spin-orbit effective field is applied the ZBP is expected to disappear as there is no topological phase. (c) The D3 device was rotated relative to the magnetic field. Starting from $0°$ where the ZBP is highest, up to $75°$, where the peak practically vanishes, and then returning back to $60°$, $50°$ and $0°$, where the peak is recovered.

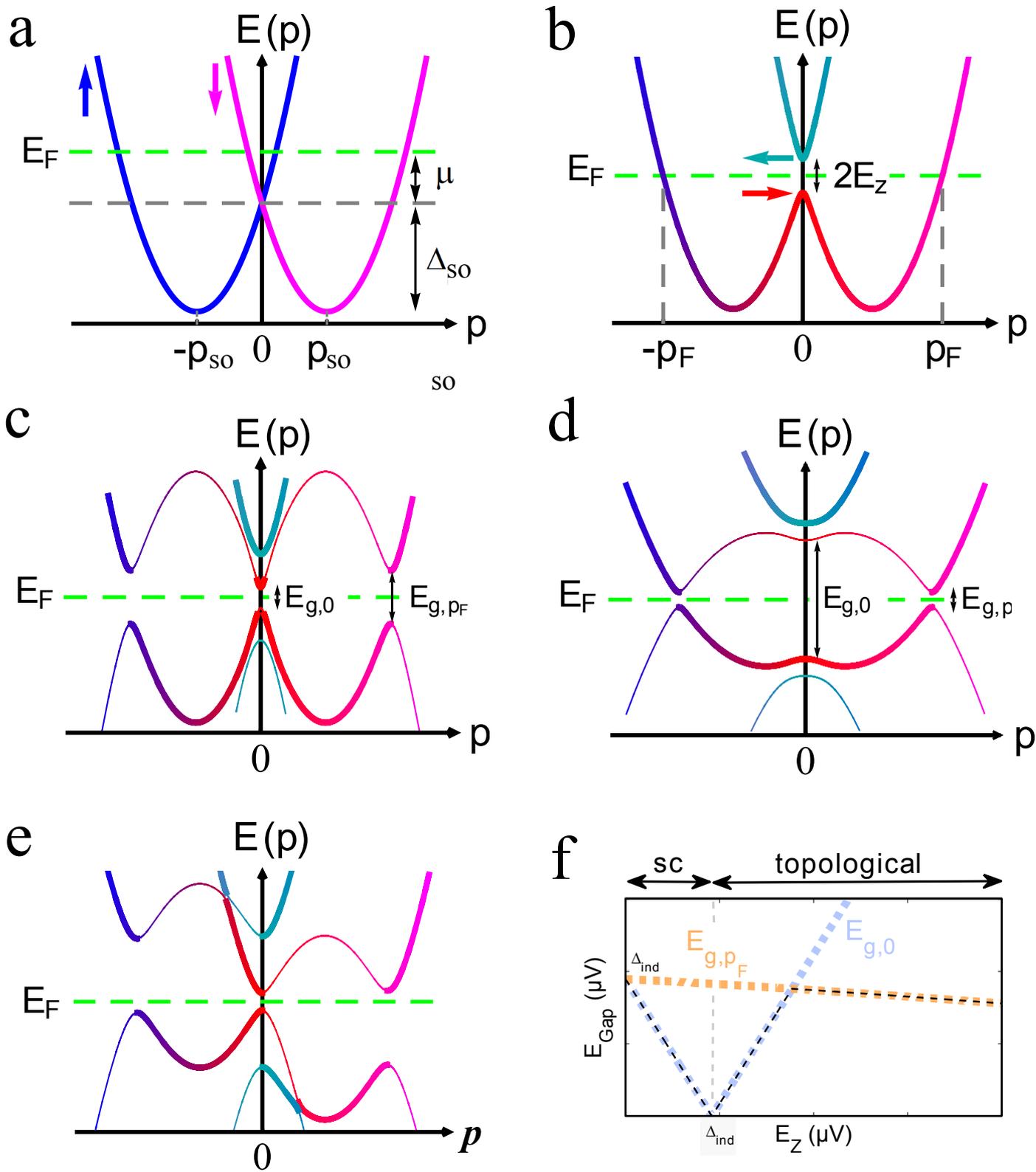

Fig. 1

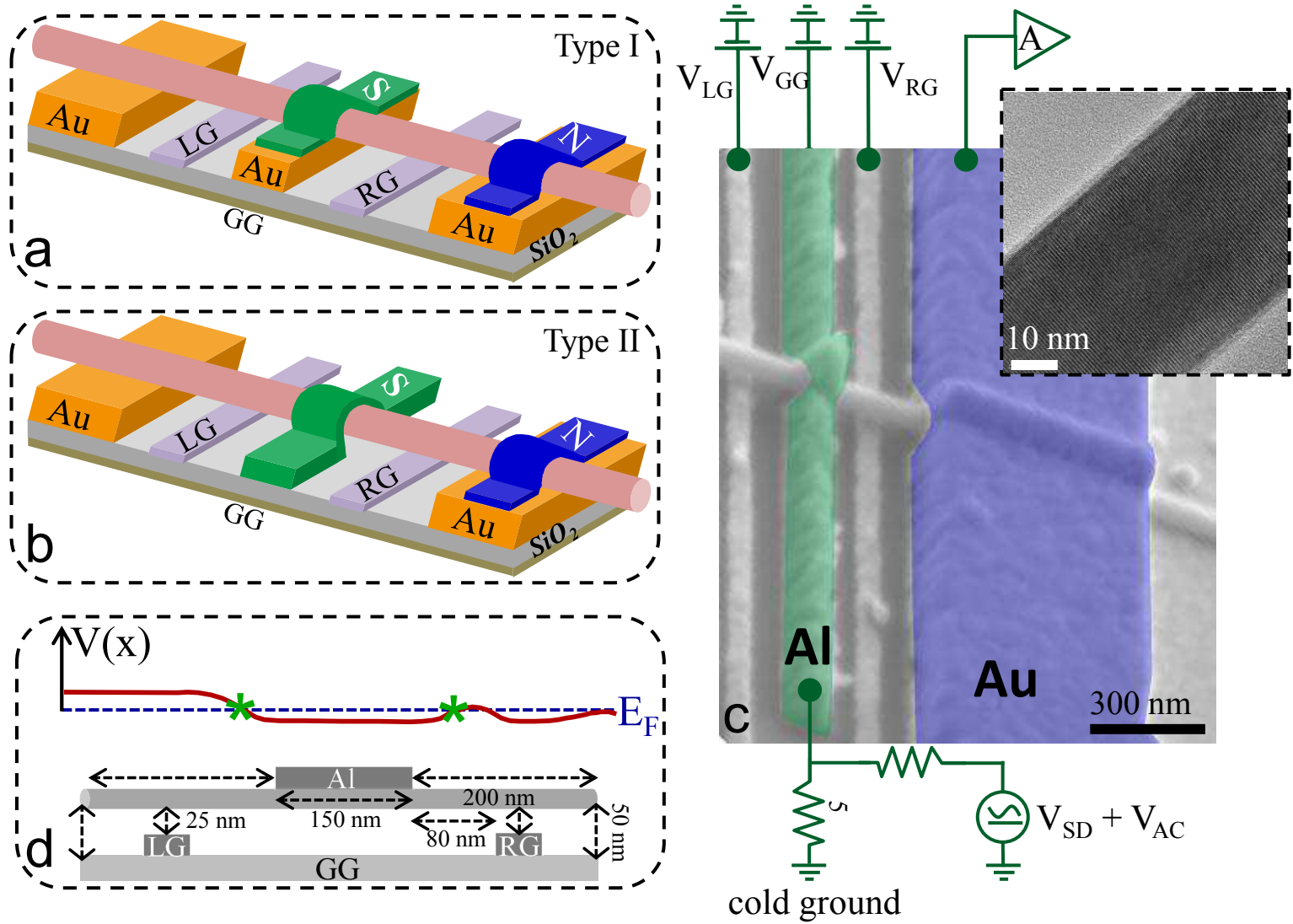

Fig. 2

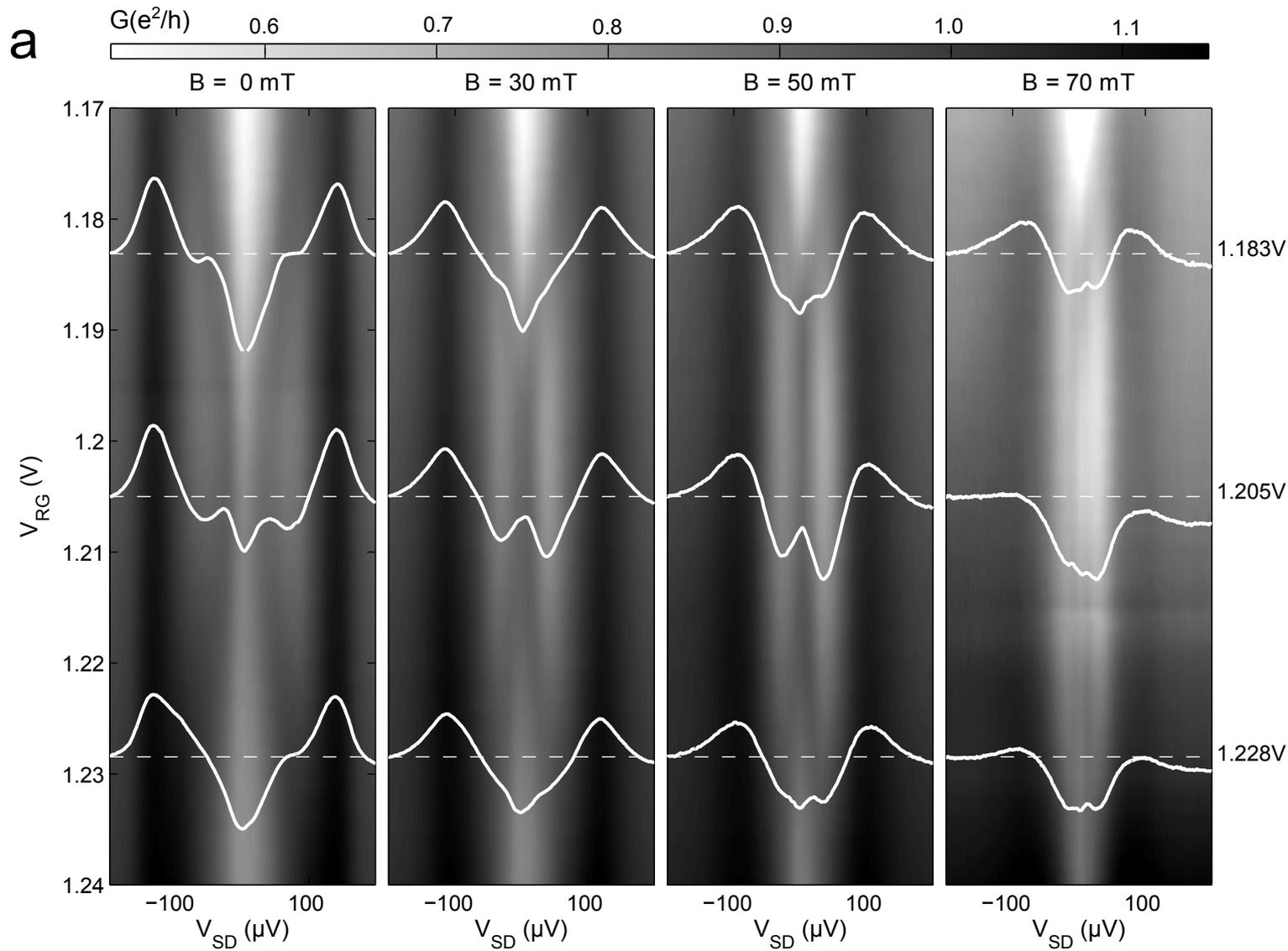

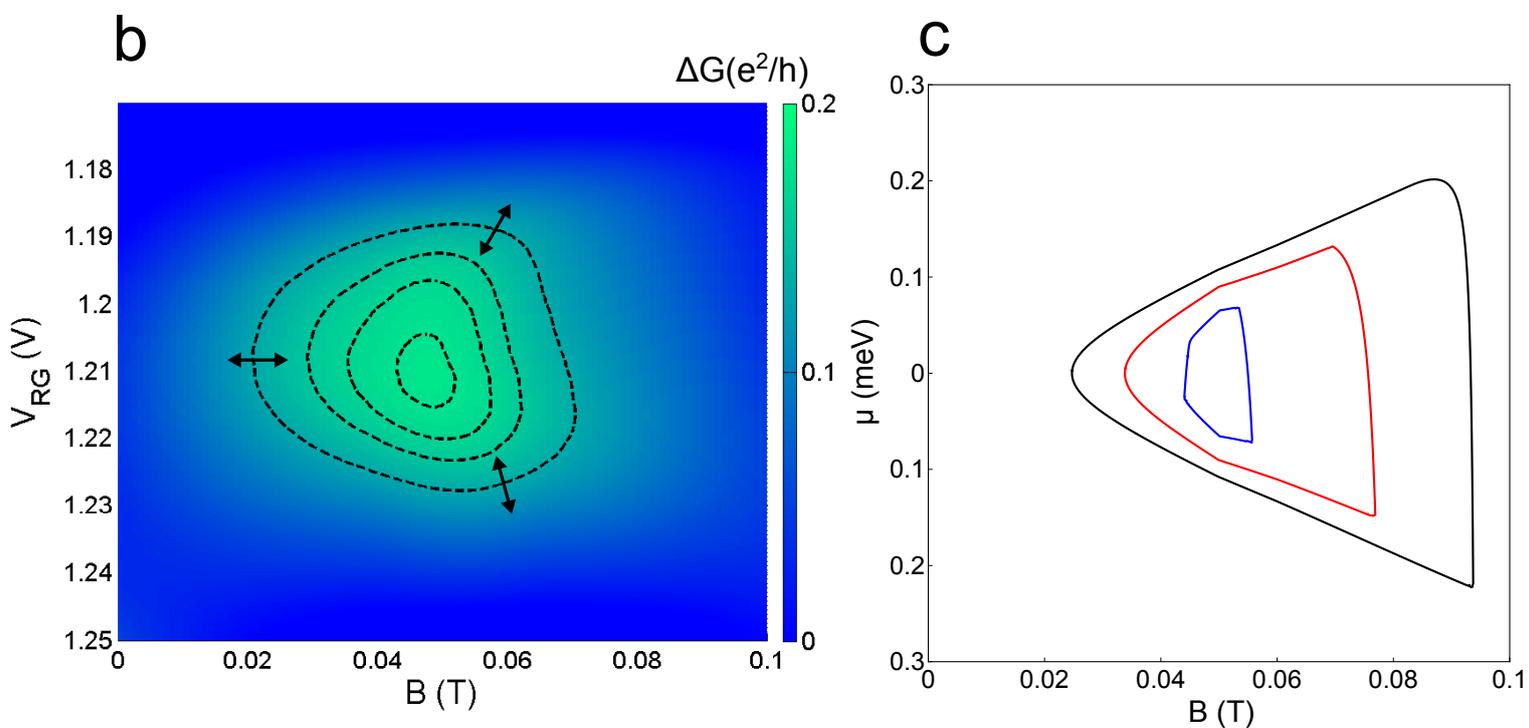

Fig. 3

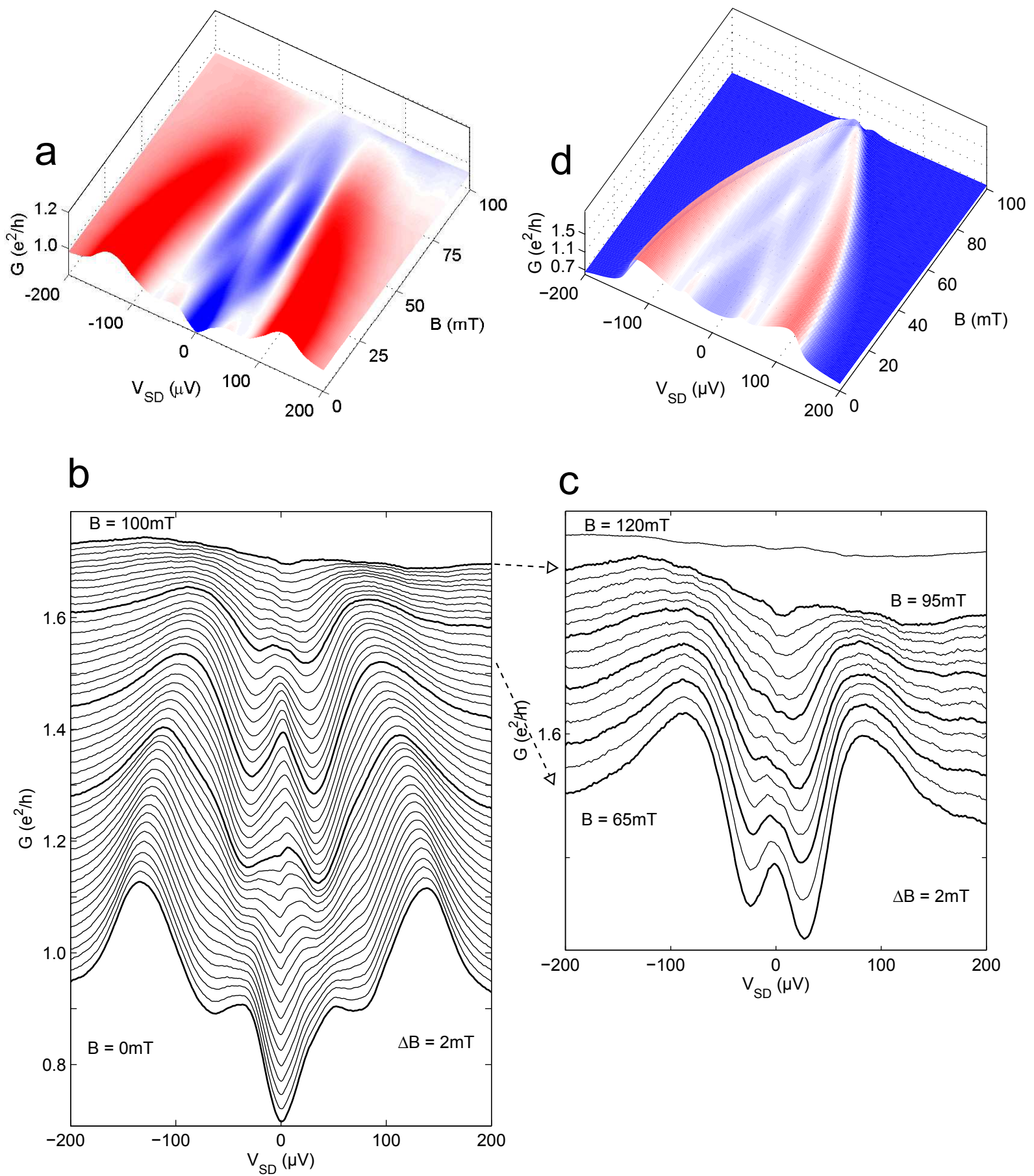

Fig. 4

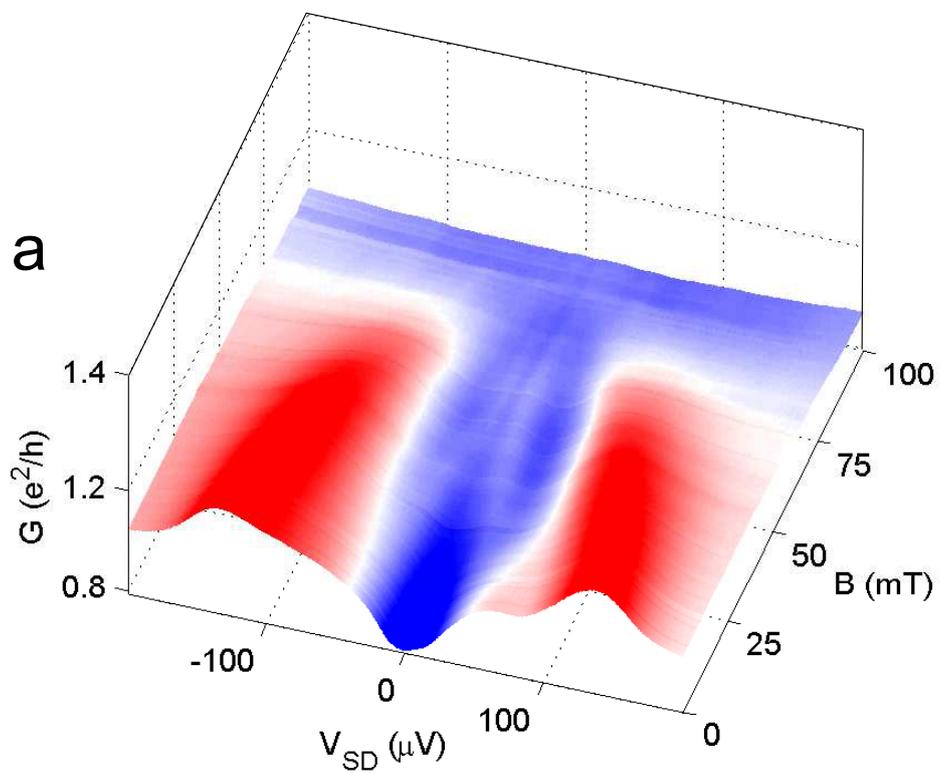

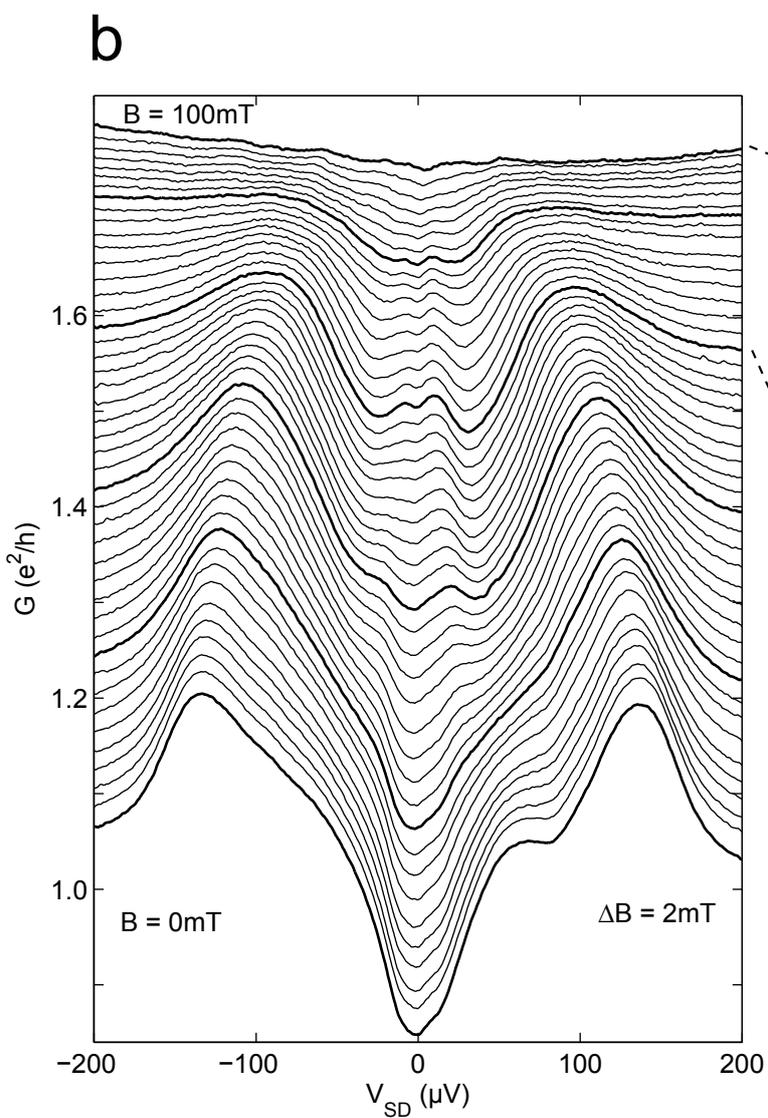
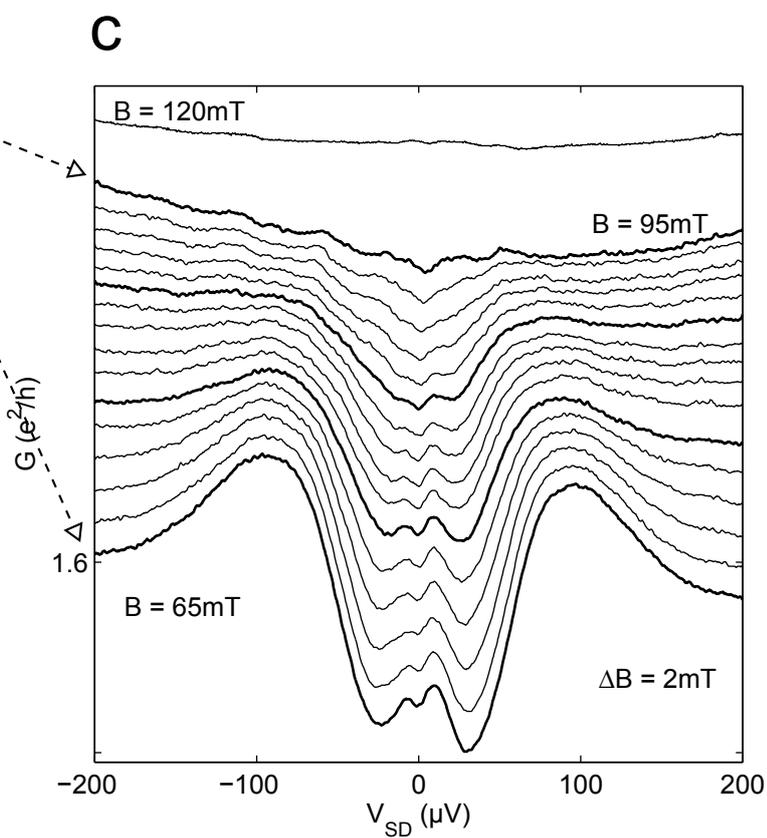

Fig. 5

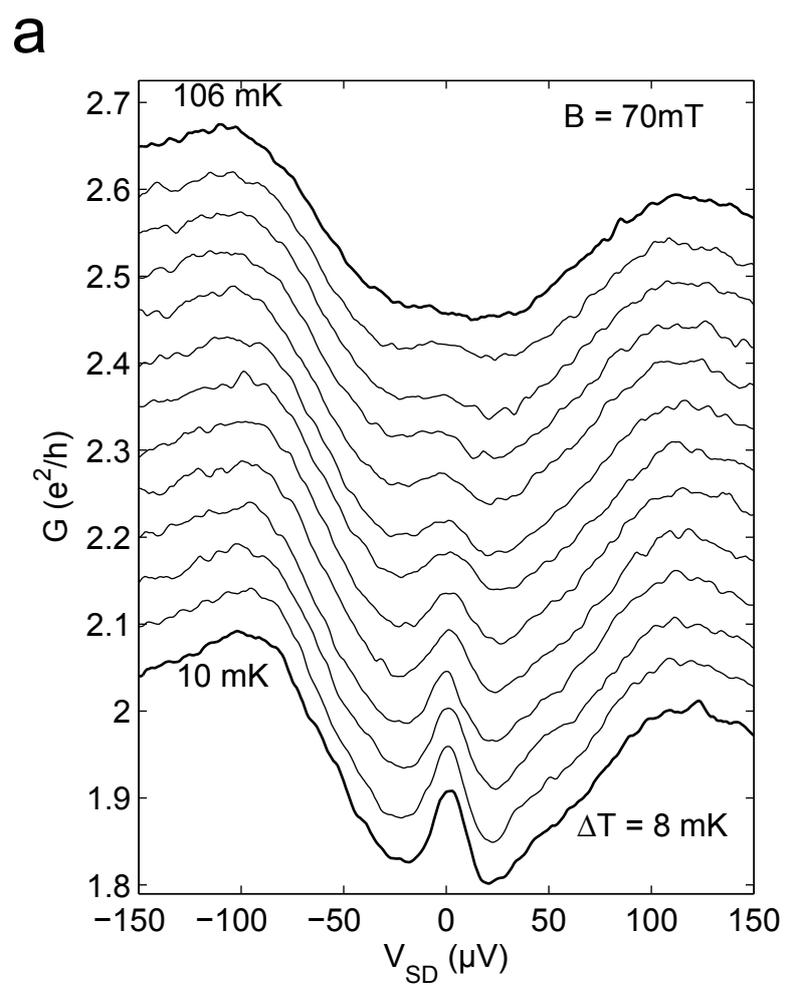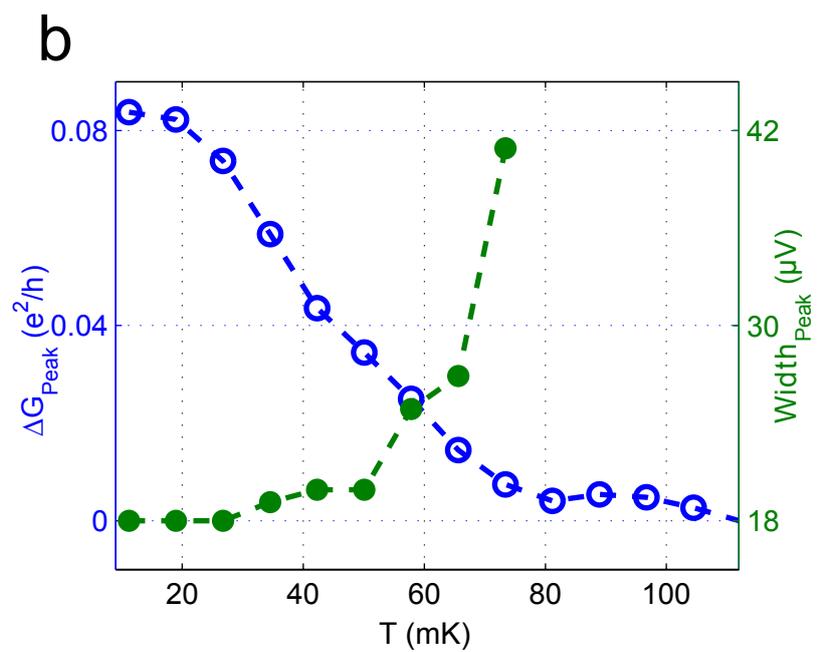

Fig. 6

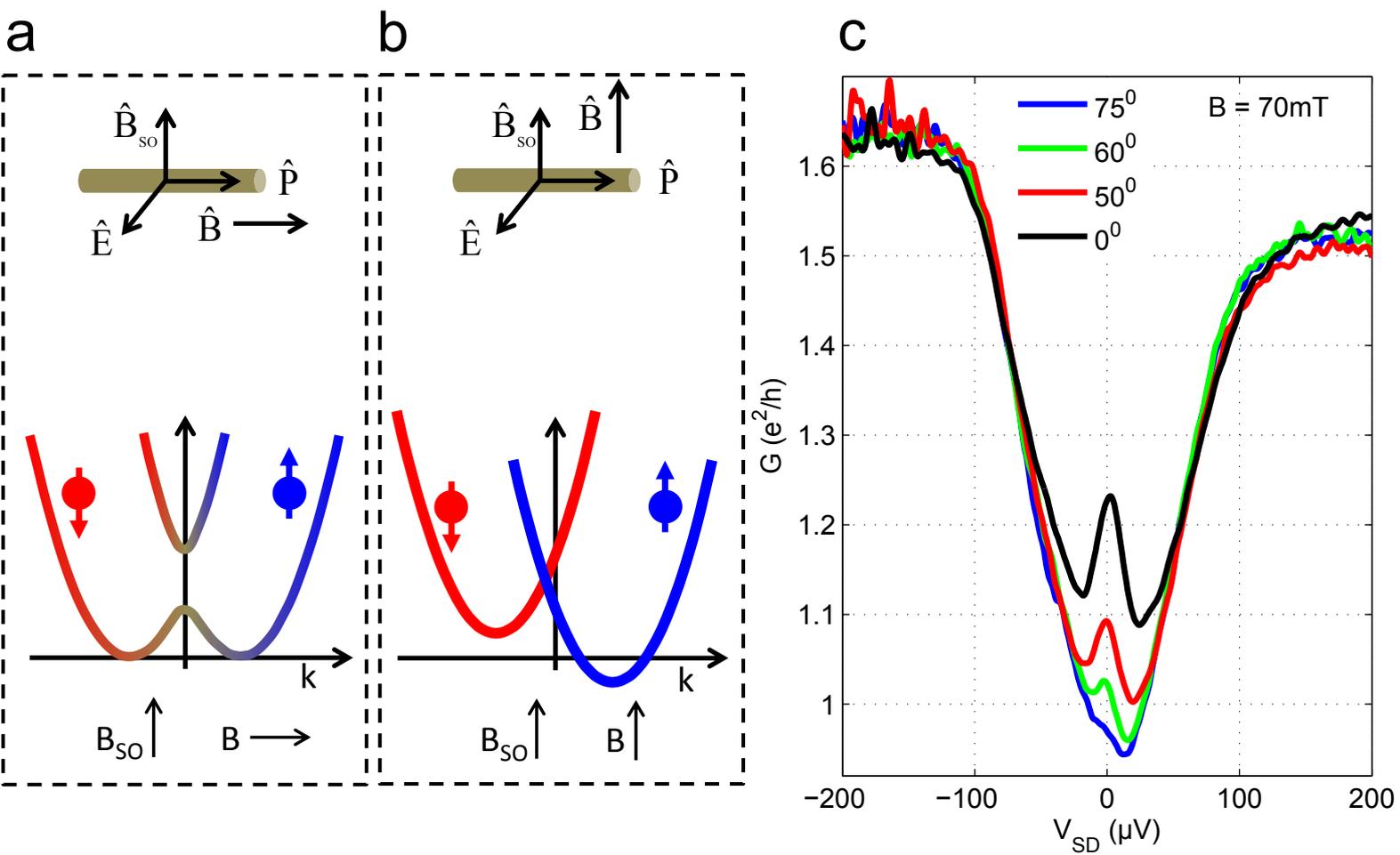

Fig. 7

## Supplementary Section

### Numerical model:

We work with a model of an infinite wire divided into $L$ segments, each segment is described by a $4 \times 4$ BdG Hamiltonian $\mathcal{H}_l(p)$ ($l = 1, \dots, L$) of the form described in Ref [1]. The Hamiltonians of the segments depend on three parameters: $\mu$, $E_z$ and $\Delta$. For example, a segment with $\Delta=0$ represents a normal wire (such as the leads of the device), and a segment with $E_z>\Delta$ and $\mu=0$ represents a topological superconductor (such a segment is used in the simulation of Fig. 4d). The total Hamiltonian density is:

$$H(x,p) = \sum_{l=1}^{L} \Pi_l(x) H_l(p),$$

where $\Pi_l(x)$ is a boxcar function which equals 1 when $x$ is inside segment $l$ and zero otherwise (segments 1 and $L$ are semi-infinite). Given an energy $E$, we find the momenta $p_{lm}^E$ (m= 1,2, ...) of the modes in each segment, as well as the corresponding eigenvectors:

$$\mathcal{H}_l(p_{lm}^E) v_{lm}^E = E v_{lm}^E.$$

After discarding divergent modes in the left-most and right-most segments, we write a wavefunction of a scattering state which is a general linear combination of the modes in each segment (the index $j$ goes over all the scattering states):

$$w_j^E(x) = \sum_{l,m} \Pi_l(x) a_{jlm}^E v_{lm}^E e^{i p_{lm}^E x}$$

We then determine the coefficients $a_{jlm}^E$ by requiring that the wavefunction and its derivative are continuous at the interfaces between the segments (we can also add barriers in the form of delta-function potentials, which will impose jumps in the wavefunction derivative). The modes with real momenta and positive (negative) group velocity in the left-most (right-most) segment are incoming modes, and it turns out that choosing the coefficients of the incoming modes determines the coefficients of the rest of the modes. We can therefore choose a basis for the scattering states such that each basis state will have only one incoming mode, and separate them into right-incoming and left-incoming states. We normalize the scattering states such that:

$$\langle w_j^E, w_k^{E'} \rangle = \delta(E - E')\delta_{jk}$$

The charge density and charge current carried by a scattering state:

$$\rho[w] = e\, w^\dagger \tau_z w$$

$$J[w] = e\, \text{Re}\left[w^\dagger \left(\frac{p}{m} + \alpha_{SO}\sigma_z\tau_z\right)w\right]$$

where $\alpha_{SO}$ is the spin-orbit constant, and $\sigma_z\tau_z$ are Pauli matrices in spin and particle-hole space just as in [1]. The charge continuity equation is:

$$\frac{\partial\rho[w]}{\partial t} + \frac{\partial J[w]}{\partial x} = 4e\Delta\,\text{Im}[w^\dagger \sigma_x \tau_x w]$$

From this equation we see that there is a source (or drain) term due to the superconductor (where $\Delta \neq 0$). This is an important property of the BTK model [2], which implicitly accounts for the current that goes through the superconductor and into the ground. To obtain the current we need to integrate over the energy, using the Fermi-Dirac distribution as the occupation:

$$I = \int dE \left\{ f_{FD}(E - eV) \sum_{j\in\{\text{left-incoming}\}} J[w_j^E] + f_{FD}(E) \sum_{j\in\{\text{right-incoming}\}} J[w_j^E] \right\}$$

Here we have attributed the entire voltage drop to the electrons incoming from the left. It is also possible to divide the voltage drop differently between electrons incoming from left and right, to better capture the real capacitance relations of the contacts. The differential conductance is:

$$G = \frac{dI}{dV} = -e \int dE\, f'_{FD}(E - eV) \sum_{j\in\{\text{left-incoming}\}} J[w_j^E]$$

To account for the appearance of the two superconducting gaps ($\Delta_{ind}$ and $\Delta_{Al}$), we have added up the conductance of two different structures. The main channel is simulated by a wire with a topological segment that has the proximity-induced pairing potential of $\Delta_{ind}$. A second channel is simulated by a wire which ends in a superconducting segment with the Aluminum pairing potential of $\Delta_{Al}$.

This two-channel technic a practical way to simulate the main features of the experiment within the limitations of the numerical model. In the experimental devices (when one of the electrodes is pinched off) all the electrons eventually go into the Aluminum, and when their energy is below the Aluminum superconducting gap, do so almost exclusively by Andreev reflections. To account for this we only take the contribution of Andreev reflection into account when calculating the conductance in the main channel.

## Experimental Section:

### InAs NWs on (011) InAs substrate by Au-Assisted MBE Growth

In agreement with the theoretical estimates and previously reported work [3], a diminishing amount of stacking faults (SFs) was observed as the NWs' diameter was reduced. Indeed, we were able to grow SF-free InAs NWs 8–10nm in diameter and 5–10$\mu$m long, having a pure wurtzite structure. Nevertheless, the real challenge lied in the growth of thicker SF-free InAs NWs. Growth conditions being crucial in determining the crystal structure, the study was carried out simultaneously on three different growth surfaces (111)B, (311)B, and (011) for which impinging group III atoms are bound to have different surface mobility. Roughness buildup of surface bulk growth in between the NWs is different on these three planes, affecting further the mobility of group III (In) atoms. Arsenic overpressure, being the parameter determining the growth rate, was kept at a V/III ratio of 100, so as to obtain a slow growth rate, and thus, nearly perfect growth. Substrate temperature was kept at 400°C, in order to favor growth of NWs over bulk deposition in between the NWs. At this low temperature, Au-assisted growth of InAs nanowires is considered to take place via the vapor–solid–solid (VSS) mechanism, since the temperature is well below the eutectic temperature of indium and gold 452°C. This is the measure we have taken in order to reduce the chances of gold incorporation during NWs growth. A high density NWs is achieved on all three surfaces, namely (111)B, (311)B, and (011). The results were described in some detail elsewhere [4]. The predominant morphology of the InAs NWs was rod shape with diameters determined by the gold droplet, ranging between 40 and 70nm, and ~5µm long. A rather small tapering, of at most 10% over 5µm, was characteristic of the InAs NWs. Interestingly, different occurrence of SFs was found in NWs having similar diameter and length, but grown on three different types of surfaces. Actually, SFs-free NWs were found only on those grown on the (011) surface. A typical example, including a TEM image and electron diffraction of a single wire, as well as a tilted SEM view of the as-grown sample, is presented in Fig. S-1. NWs for this study were taken solely from this type of samples.

**Sample fabrication and measurement**

The sample was fabricated on a thermally oxidized *Si*. The nanowire was suspended on two gold pillars, 50nm high. Two lower gold pillars, 25nm high, provided local gating on both sides of the wire. After the wires were spread from ethanol solution, the source and drain regions were etched by ammonium polysulphide $(NH_4)_2S_x$=1:5M), in order to remove the native oxide, and immediately transferred into the evaporation chamber. For 'normal' metallic contacts 5/100nm Ti/Au were evaporated, while for the superconducting contact 5/100nm Ti/Al was evaporated.

The superconducting contact was biased by a voltage source, DC or AC (a voltage divider was placed on the cold finger); with an AC excitation voltage ~ 1-2µV. Conductance measurements were conducted at low frequency using room temperature current preamplifier at 575Hz with gain $10^7$, input impedance ~200 ohm and current noise ~50 fA/$\sqrt{Hz}$. We also measured the conductance at a higher frequency (~1MHz), employing a low noise voltage cold preamplifier cooled to 1K (see another work [5]).

**Conductance measurement**

Fig. S2 shows a typical response of the devices as a function of local gate voltage for a fixed $V_{GG}$. In both the devices conductance oscillates due to Febry-Perot (FB) type fluctuations. It can be seen from the figure that at zero gate voltage both devices are conductive (n-type), however, in type I (D1) the 2nd sub-band is populated, whereas at type II (D3) only the 1st sub-band is populated. Note, that the conductance exceeds $2e^2/h$ in the 1st sub-band due to Andreev reflection.

**Noise measurement**

Current is injected from the Al superconductor carrying fluctuations, shot noise [5]. Shot noise depends linearly on the current $I$ and on the tunneling charge $e^*$. The 'low frequency' spectral density of the 'excess noise' (shot noise above the Johnson-Nyquist and environment noise) in the single InAs channel takes the form: $S_i(0)=2e^*I(1-t^*)F(T)$, with $t^*=t$ for electrons and $t^*=t^2$ for Cooper pairs, and $F(T)=coth\xi-1/\xi$, with $\xi=e^*V_{SD}/k_BT$. In Fig. S3b we plot $S_i(0)$ as a function of $I$ for zero magnetic field $B$ and for $B\sim0.2T$ for the type I device (D1). The blue and red solid lines are the 10mK predictions for $t^*=0.41$, $e^*=2e$ and $t=0.63$, $e^*=e$, respectively; demonstrating an excellent quantitative agreement with the data (blue circles). The distinct change of slope (from $e^*=2e$ to $e^*=e$) nicely corresponds to $\Delta$ of the Al (S3a). A perpendicular small magnetic field (~0.2T) quenched the superconductivity with the excess noise nicely agreeing with $e^*=e$ across the full biasing range. An essential fact regarding the numbers of occupied bands in the wire is achieved by the observing that the fit to the shot noise depends on the t* which assumes single band occupied.

**Superconducting gap**

In fig. S4 the dependence of $2\Delta_{Al}$ as a function of B in type II device (D3) is shown. There is no appearance of ZBP with magnetic field, and the critical field can be clearly seen more than 150mT.

**'Reflection-less tunneling' and Andreev bound state**

Constructive interference between electron reflection and Andreev reflection in S-I-N-I devices (where I - insulator, S - superconductor, N - normal), which is known as 'Reflection-less Tunneling' is shown in Fig. S5a for the type I device (D2). As expected, the ZBP diminishes with increasing B. Fig. S5b shows the B dependent Andreev bound state of the type II device (D3). From the splitting we the Lande g-factor ~20 was extracted. In Fig. S5c we have shown we performed a simulation of an S-I-N-I-N device at 30mK.

**Kondo in Coulomb blockade regime.**

Fig. S6 shows the Coulomb blockade oscillations with Kondo signature of the type I device (D2) at B = 0T. 2D gray plot with bias and local gate voltage clearly shows the Kondo enhanced ZBP at B = 0T with odd number of occupied electrons. As expected, Kondo exhibits Zeeman splitting with B. From the Kondo splitting the Lande g-factor was extracted to be ~13.

**Kondo-like peak in FB regime.**

In Fig S7 we observe the appearance of a broad (~100µV) ZBP after killing the superconductivity and persist up to large magnetic field for type I (D2) and type II (D4) device.

**Stability of ZBP with $V_{RG}$ and B**

Fig. S8a shows zero bias conductance of the type II device (D3) as a function of $V_{RG}$ at zero and finite B. In Fig3 of the manuscript we have shown the evolution of the ZBP (D4) in a few different magnetic fields as function of $V_{RG}$. Fig. S8b shows the data for all the B. Discussion has been included in the figure caption.

**Stability of ZBP with $V_{GG}$ and $V_{RG}$**

Fig. S9 shows the ZBP at zero bias and at 80mT as a function of $V_{RG}$ for different $V_{GG}$. It can be seen from the figure that the zero energy peak appears, and stays at zero energy, in the range of $\Delta V_{RG}$~0.2V and $\Delta V_{GG}$~1V.

**Split ZBP**

In Fig. 5 we have shown a ZBP that remains split at a wide range B for upper end of chemical potential. Fig. S10 shows that the same effect happens for the lower end of chemical potential.

**Closing and reopening of gap**

In Fig. 4 we have shown the evolution of two shoulder peaks with B at mid chemical potential. Fig. S11 shows it for another set of parameters. Details have been discussed in the figure caption.

**ZBP with B**

Fig. S12 shows ZBP the in different devices where we see the emergent of the ZBP at finite B which disappears with the killing of superconductivity. However, we do not see the closing of shoulder peaks at small B like in Fig4 in the manuscript.

**Figure Captions:**

**Figure S1**. (a) High resolution transmission electron microscope image of a typical InAs nanowire taken from the sample used in this study. Image is taken from the <0011> zone axis and depicts a pure wurtzite structure and a growth direction along the <0001> C

axis. Scale bar relates to 10 nm. The inset shows the electon diffraction taken at the center of the NW seen in (a) exhibiting a clean wurtzite structure and a growth direction along the <0001> C axis marked by an arrow. TEM image is courtesy of Ronit Popovitz-Biro. (b) Scanning electron microscope image of the as grown InAs nanowires taken at 45 degrees. Nanowires were grown in the <111> direction by Au-assisted vapor-liquid-solid (VLS) molecular beam epitaxy (MBE) on a (011) InAs substrate. Scale bar relates to 1 micron.

**Figure S2.** (a) $G_R$ vs $V_{RG}$ for the type I device (D1) at $V_{GG} = 0$V. The dashed red lines clearly indicate the $1^{st}$ and $2^{nd}$ sub-band, respectively. (b) $G_R$ vs $V_{RG}$ for the type II device (D3) at $V_{GG} = 0$V, where $1^{st}$ sub-band is indicated by the dashed red line. The conductance oscillates due to Febry-Perot fluctuations.

**Figure S3**. Non-linear conductance (top) and auto correlation signal (shot noise, bottom) as a function of current, Isd, for $B$=0 and $B$=0.2T in type I device (D1). Solid lines are theoretical predictions at $T$=10mK (assuming 1 occupied band). Charge is $2e$ for $V_{SD}<\Delta$ (blue line) and $e$ for $V_{SD}>\Delta$ (red line).

**Figure S4**. (a) The color scale of $2\Delta_{Al}$ as a function of B of the type II device (D3) without a ZBP. (b) Cuts for every $\Delta B = 4$mT. The critical field can be clearly seen more than 150mT.

**Figure S5**. (a) ZBP at B = 0T due to 'reflection less tunneling' and its dependence with B of the type I device (D2). The ZBP diminishes with increasing B. (b) ZBP at B = 0T due to ABS and its splitting with B of the type II device (D3). The voltage drop across the sample was assumed to drop equally on the two potential barriers at the interfaces of the Al superconductor and the gold 'normal' contact. The fitted Lande g-factor is ~20. (c) A simulation of an S-I-N-I-N device, where the middle N section is of 150nm length and the insulators are delta-function barriers of 4.5Vnm strength. A second channel was added modeling the direct tunneling into the Al superconductor, with barrier strength of 0.4V-nm. The fitted g-factor from the experiment was used, as well as the measured dependence of $\Delta_{Al}$ and $\Delta_{ind}$ on the magnetic field. The data was convolved with a Fermi-Dirac kernel to simulate a temperature of 30mK.

**Figure S6.** (a) Coulomb blockade oscillations with Kondo signature of the type I device (D2) at B = 0T. (b) 2D gray plot with bias and local gate voltage clearly shows the Kondo enhanced ZBP at B = 0T with odd number of occupied electrons (c) 2D gray plot with splitted Kondo peaks at B = 30mT. (d) and (e) are the cuts at $V_{RG}$ = -0.585V for B = 0 and 30mT, respectively, as shown by the vertical dashed lines. From the Kondo splitting we calculate the Lande g-factor is ~13.

**Figure S7.** (a) 2D color plot with bias and magnetic field of the type I (D2) and (b) the type II (D4), where a broad (~100µV) ZBP appears after killing the superconductivity and persist up to large magnetic field.

**Figure S8.** (a) Zero bias conductance of the type II device (D3) as a function of $V_{RG}$ at $V_{GG}$ = -13.5V for B = 0 (red line) and 50 mT (blue line). It can be seen that in the range of $V_{RG}$ from1.8 to 2V there is an enhancement in conductance at 50mT compared zero B, which indicates having ZBP with magnetic field. (b) Similar features have been observed for another type II device (D4) at $V_{GG}$ = -18.3V for $V_{RG}$ range from1.17 to 1.24V (as reported in Fig. 3 of the manuscript). From left to right panel the 2D gray plots show the evolution of ZBP in different magnetic field as function of $V_{RG}$, which controls the barrier height, however, it also affects the chemical potential under the superconductor. The cuts are taken at three local gate voltages (1.83, 1.205 and 1.228V). At B = 0T (upper left panel) one can clearly see that bulk superconducting Al gap ($\Delta_{Al}$ ~ ±150µV) with two shoulders as induced gap, $\Delta_{ind}$ ~ ±45 µV. Applying B=30mT, the two conductance shoulders come closer near the $V_{RG}$ = 1.205V and turn into a relatively wide, barely split, ZBP, which splits at higher and lower gate voltages. At B=50mT (bottom left panel), the ZBP was sharper and a more distinct single peak; persisting for a wider range of gate voltage even though remain splitted at the lowest and highest $V_{RG}$. This is likely to be a direct result of the increasing chemical potential, which lowers the gap, $E_g(0)$, and increase the coherence length of the Majorana, thus enhancing coupling and splitting. For B=70mT, the single peak splits and remains for wide range of gate voltage. This is also consequence of overlapping between two Majorana; at higher magnetic field the gap is determined by $E_g(P_F)$ which decreases with B. At B=80mT (bottom right panel) the conductance features become weaker as it approaching to the critical field ~ 100mT of the aluminum superconductor.

**Figure S9.** Stability of the ZBP with local gate as well as global gate of the type II device (D3, critical filed > 150mT). Gray scale representation of the conductance with cuts at a certain $V_{RG}$. (a) At B=0, there is a zero energy conductance dip, which results from the conductance (weak) peaks on both sides due to the induced gap as seen in D4 device. The ubiquitous Al superconducting gap is pronounced. (b) At B=80mT, the zero energy peak is nailed at $V_{SD}$=0 as function of $V_{RG}$. At the lowest and highest $V_{RG}$, a clear split in the peak is visible as seen in D4 device. (c) Another stability diagram for four different $V_{GG}$'s, showing the zero energy peak height for different gate voltages. While the peak in these voltages range stays hooked to $V_{SD}$=0, it is the highest at a more confined range of parameters. (d) A simulation of changing chemical potential, using the same parameters in the manuscript. The splitting of the peak in Fig. S9b is reconstructed.

**Figure S10.** 2D color plot with bias and magnetic field for $V_{RG}$=1.185V (lower end of chemical potential of gap) of Fig.3a in the manuscript, where the color scale figure (a) and the corresponding cuts (b) show a split peak in all regions of the magnetic field, terminated by the collapse of the superconducting Al gap, which is very similar to Fig. 5 of the manuscript.

**Figure S11.** (a) 2D color plot with bias and magnetic field for $V_{RG}$=1.21V (near the middle of chemical potential of gap) of Fig.3a in the manuscript for D4 device. The two shoulder peaks (being $\Delta_{ind}$~45μeV at B=0) move closer in energy, followed by a ZBP that emerges at B~35mT, later to split beyond B~70mT, clearly can be seen in the cuts (b)&(c), where (c) is the zoom between 65mT to 95mT. It can be clearly seen that 120mT there is complete collapse of the superconducting state.

**Figure S12.** (a) 2D color plot and (b) cuts with bias and magnetic field of the type II device (D3). ZBP appears at ~ 30mT, maximize at ~ 70 mT and persist up to ~110mT, then diminishes with increasing magnetic field. Two little superconducting shoulders can be clearly seen at 150mT. However, we do not see the closing of gap and reopening in this device. Similarly, (c) 2D color plot and (d) cuts with bias and magnetic field of the type II device (D4) in different cool down. Clearly ZBP disappears before superconductivity dies.

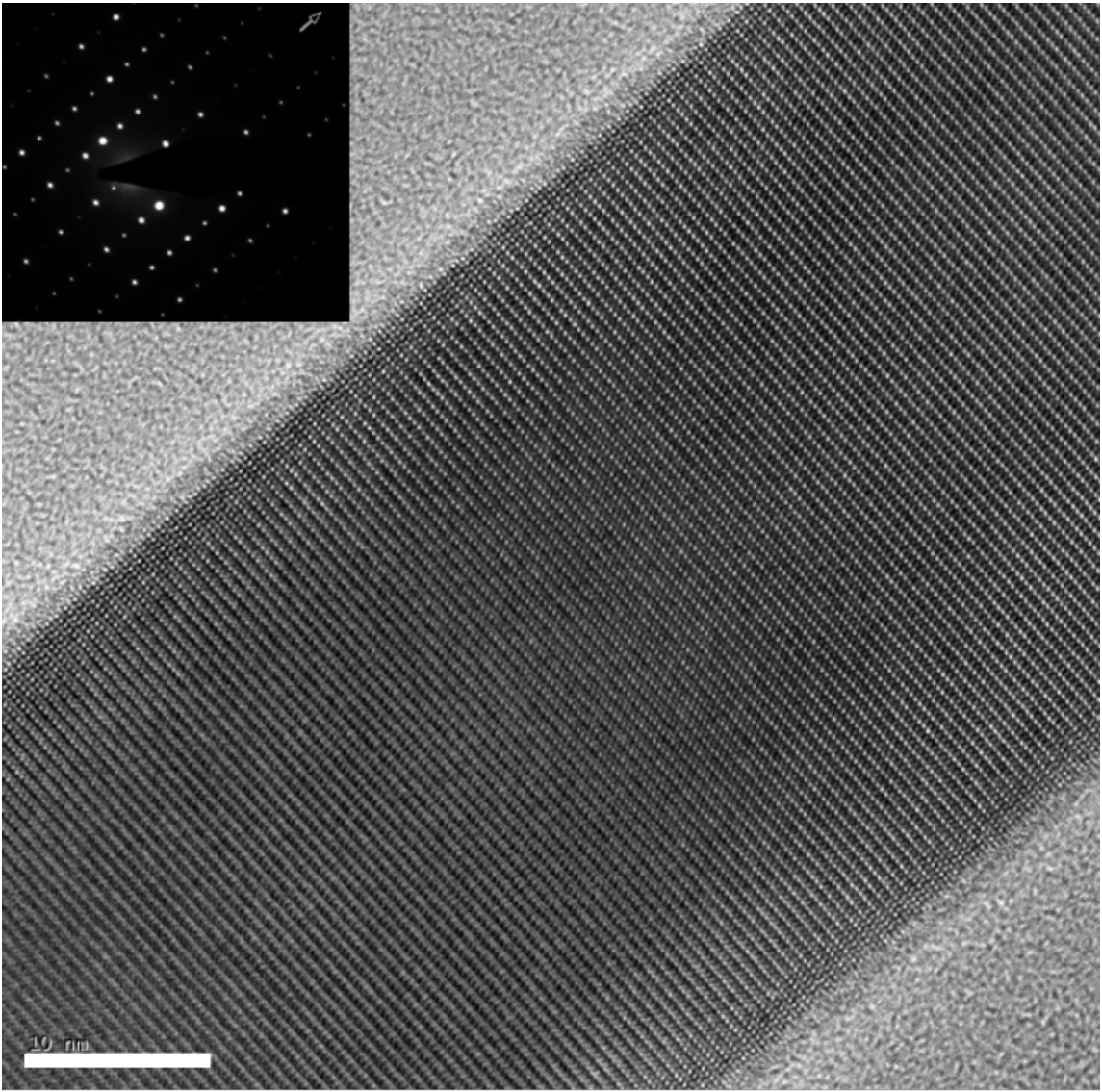
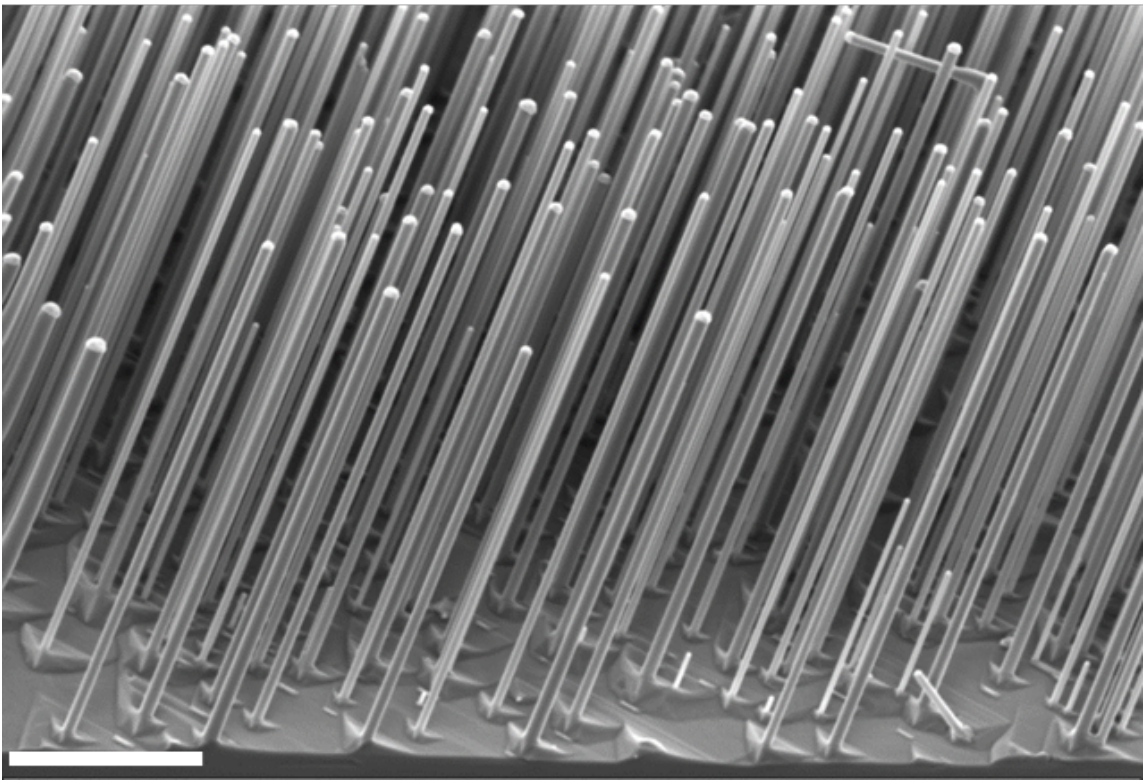

S1

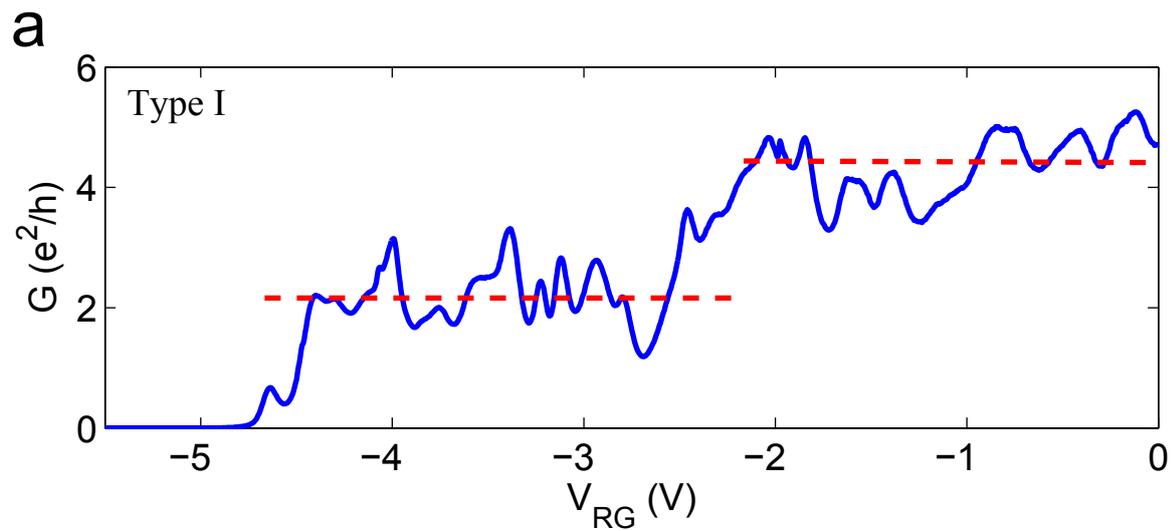
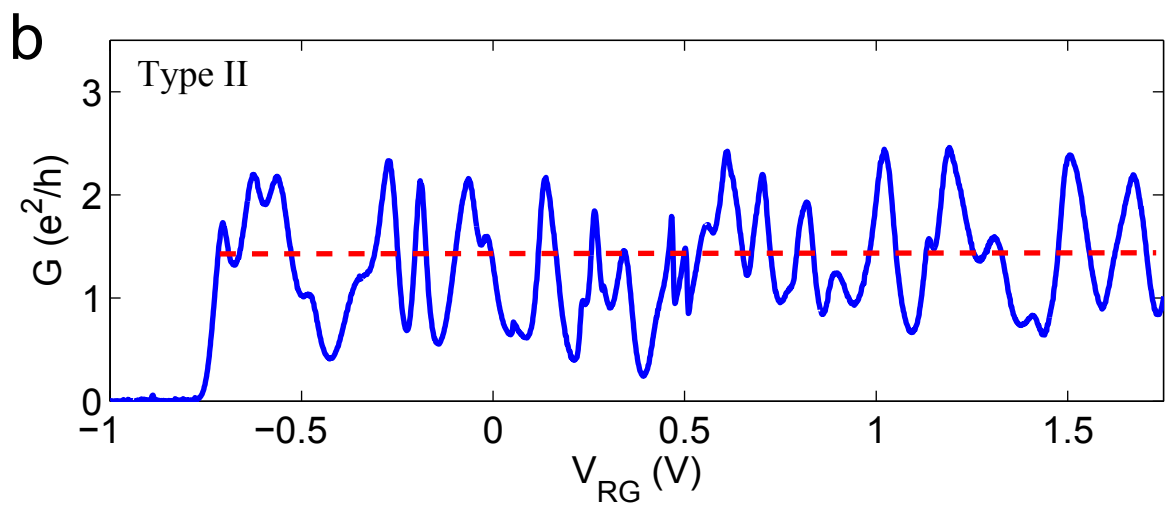

S2

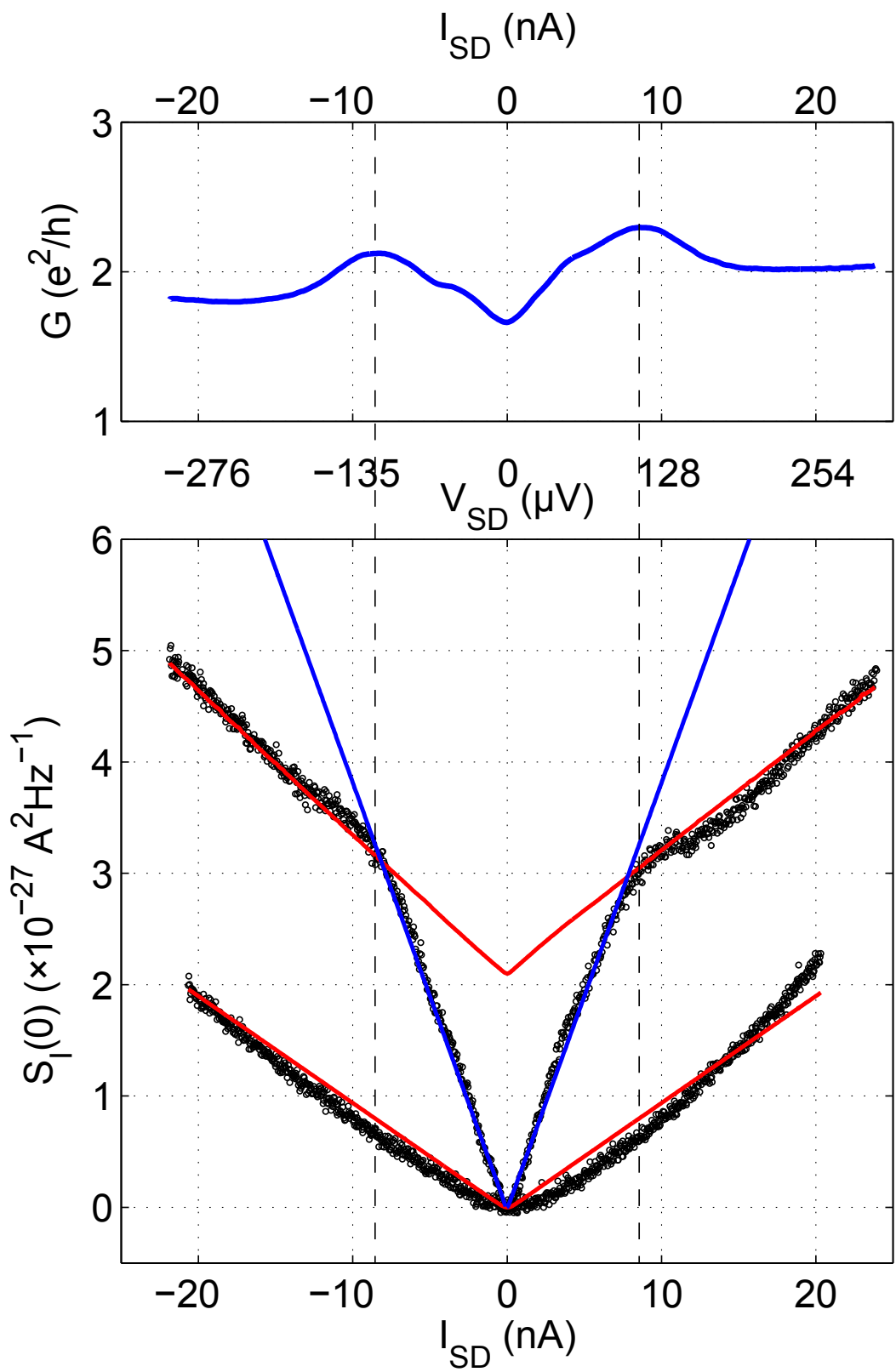

S3

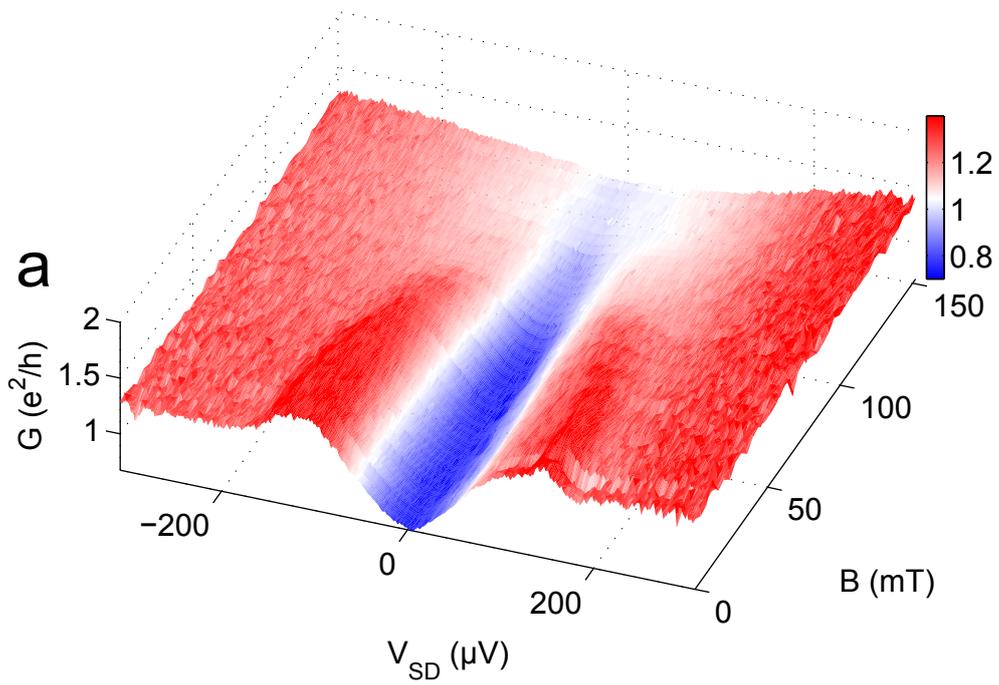

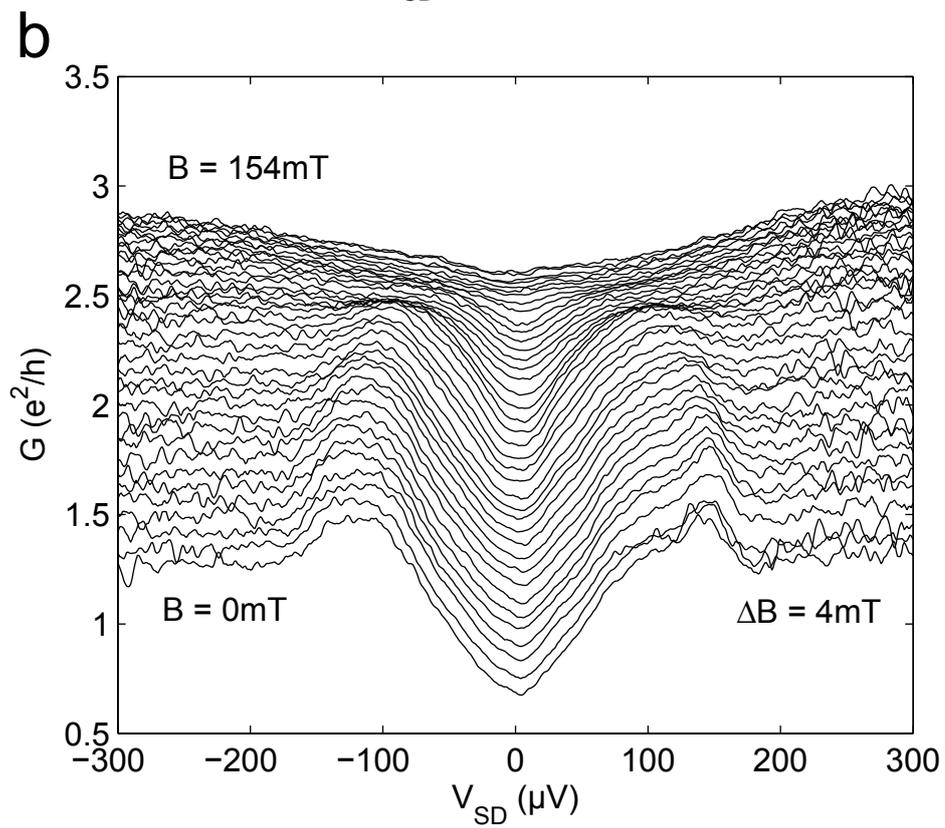

S4

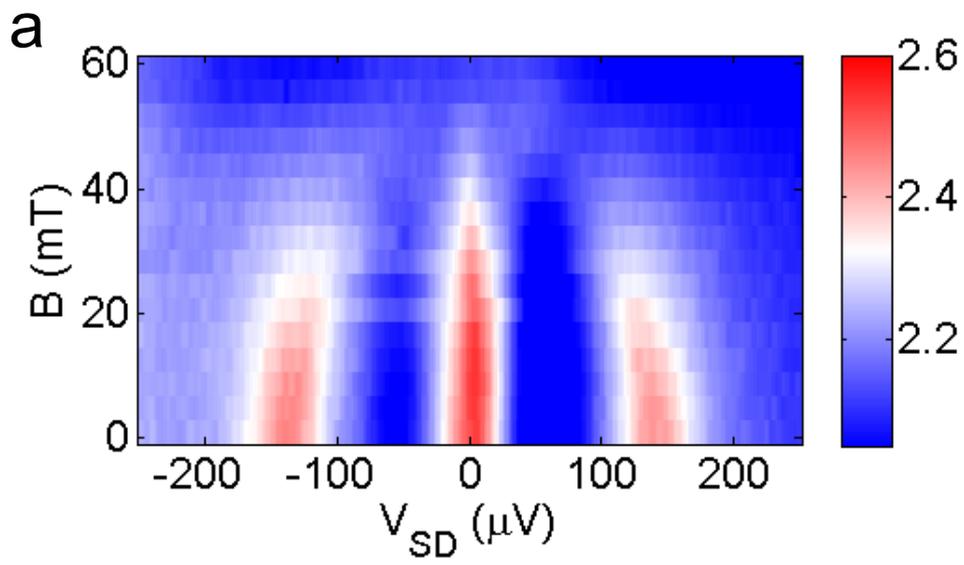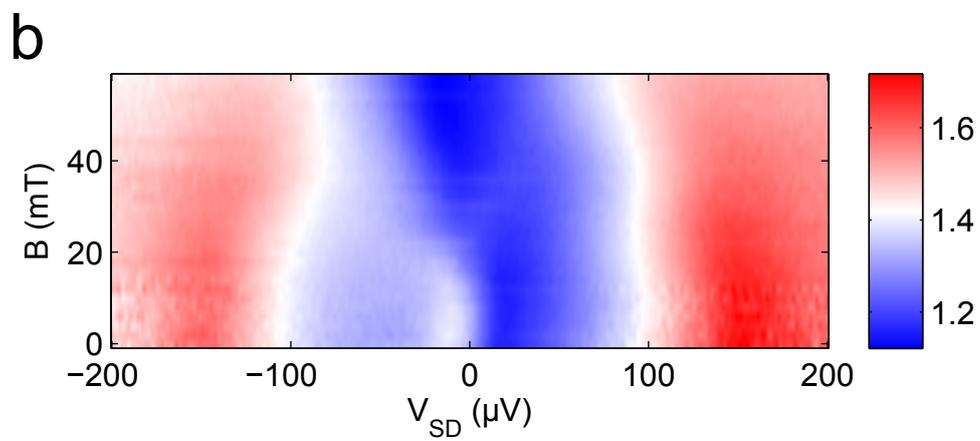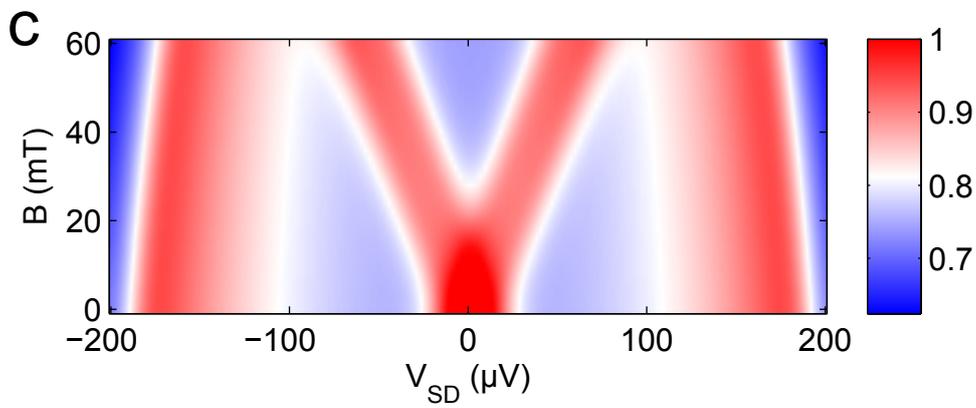

S5

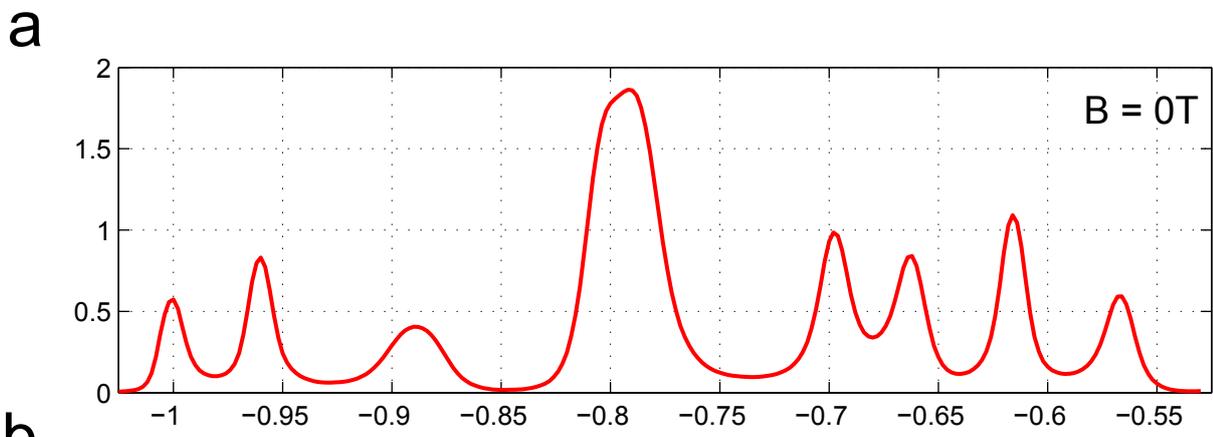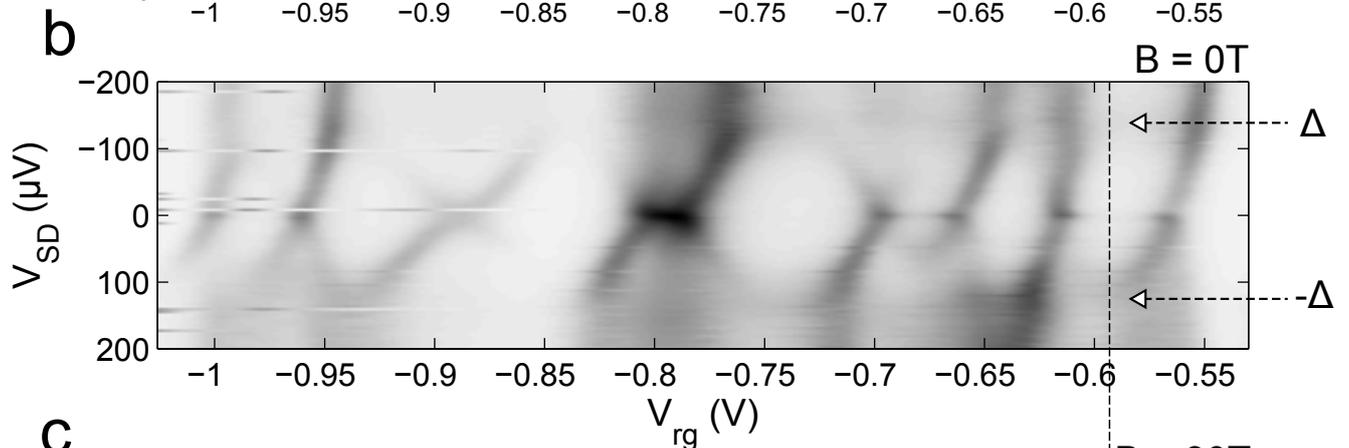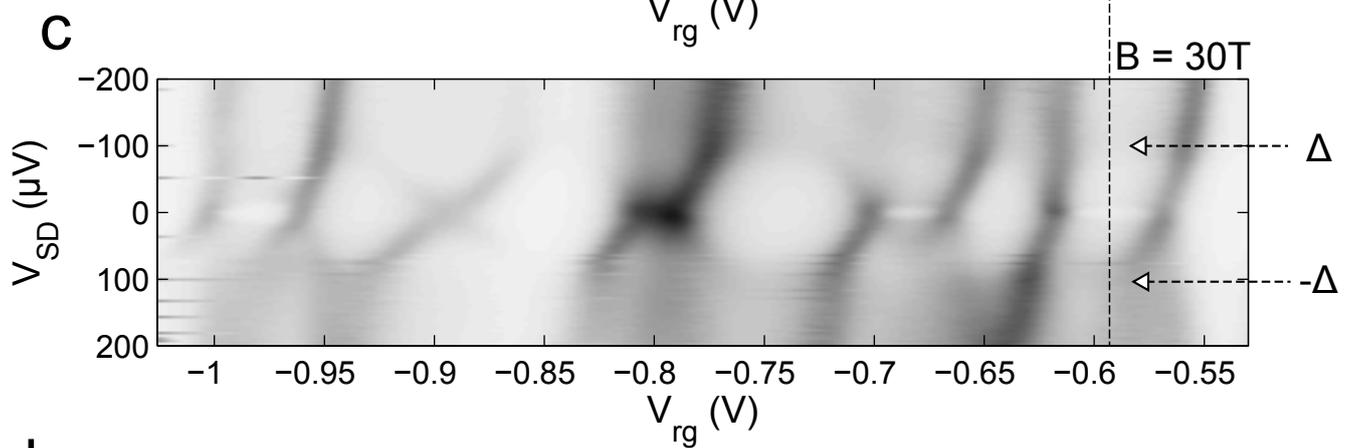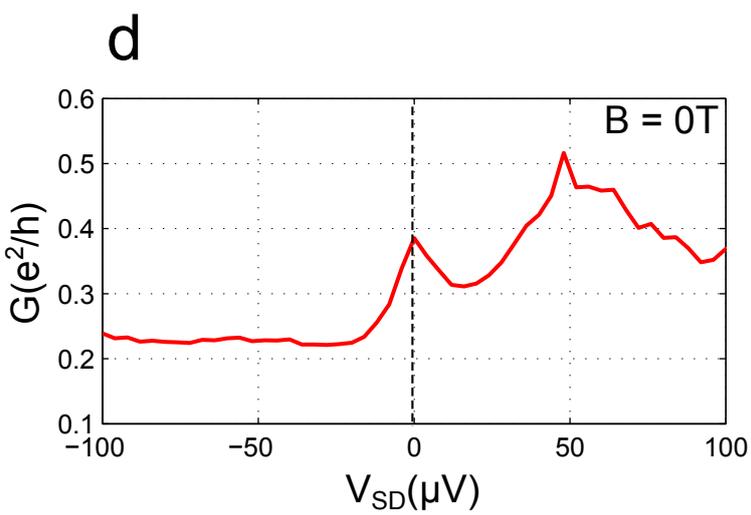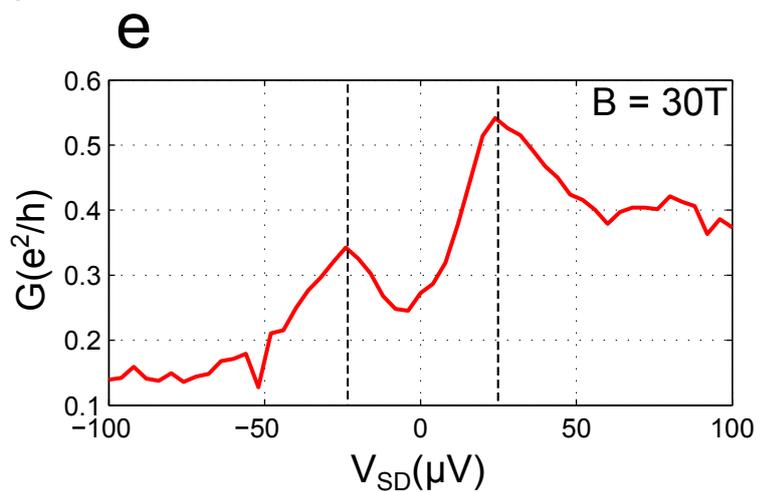

S6

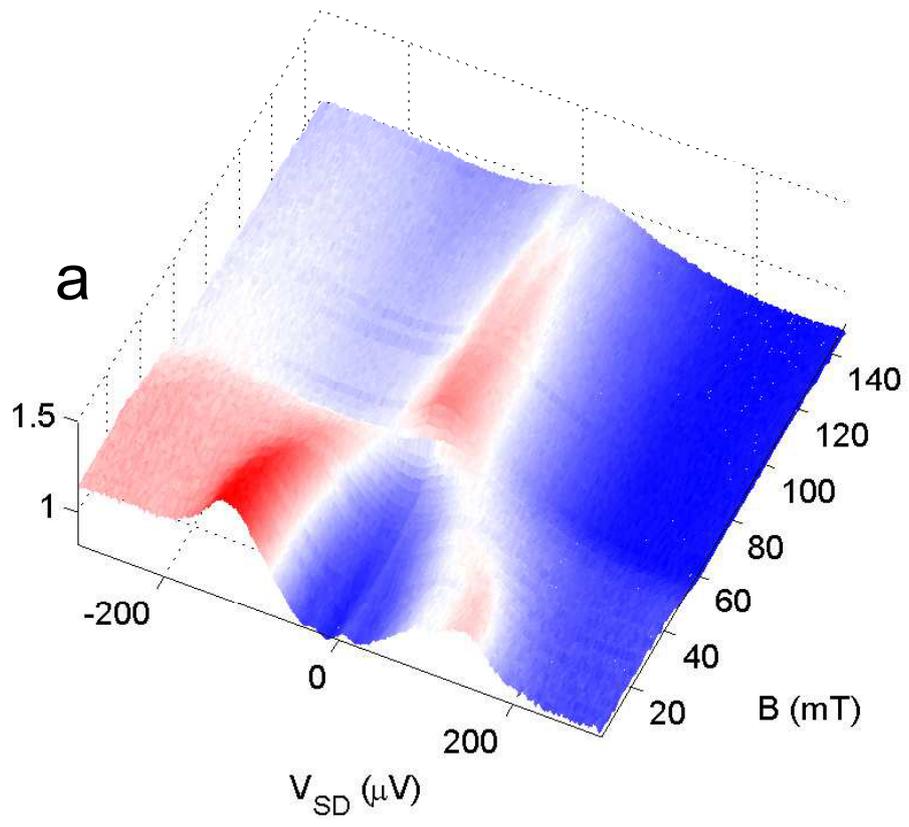
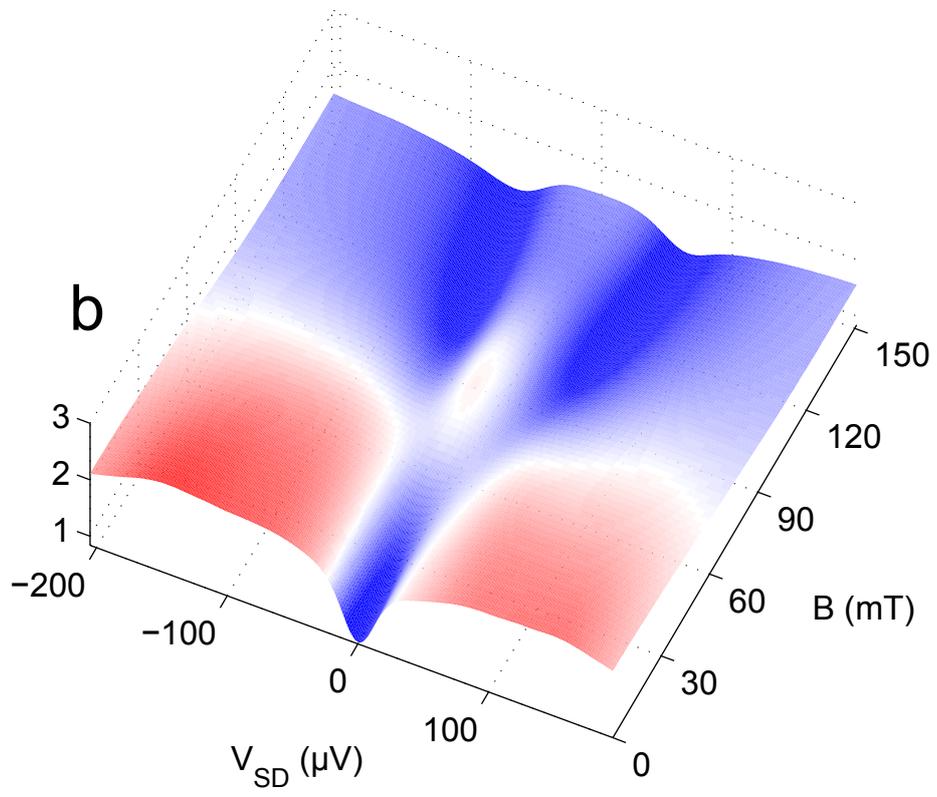

S7

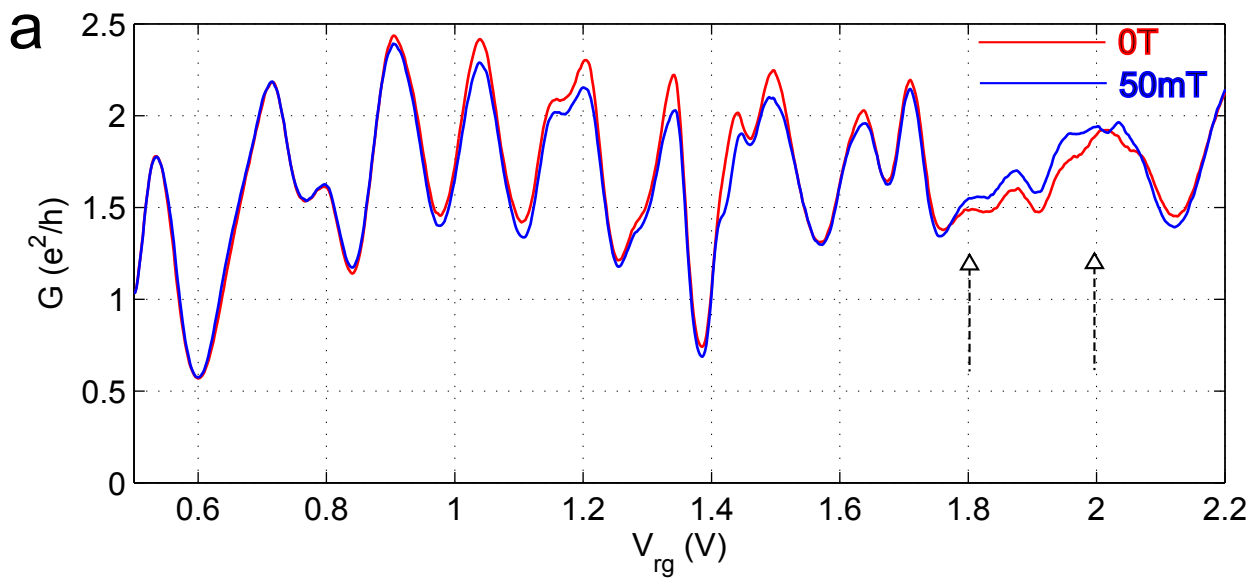

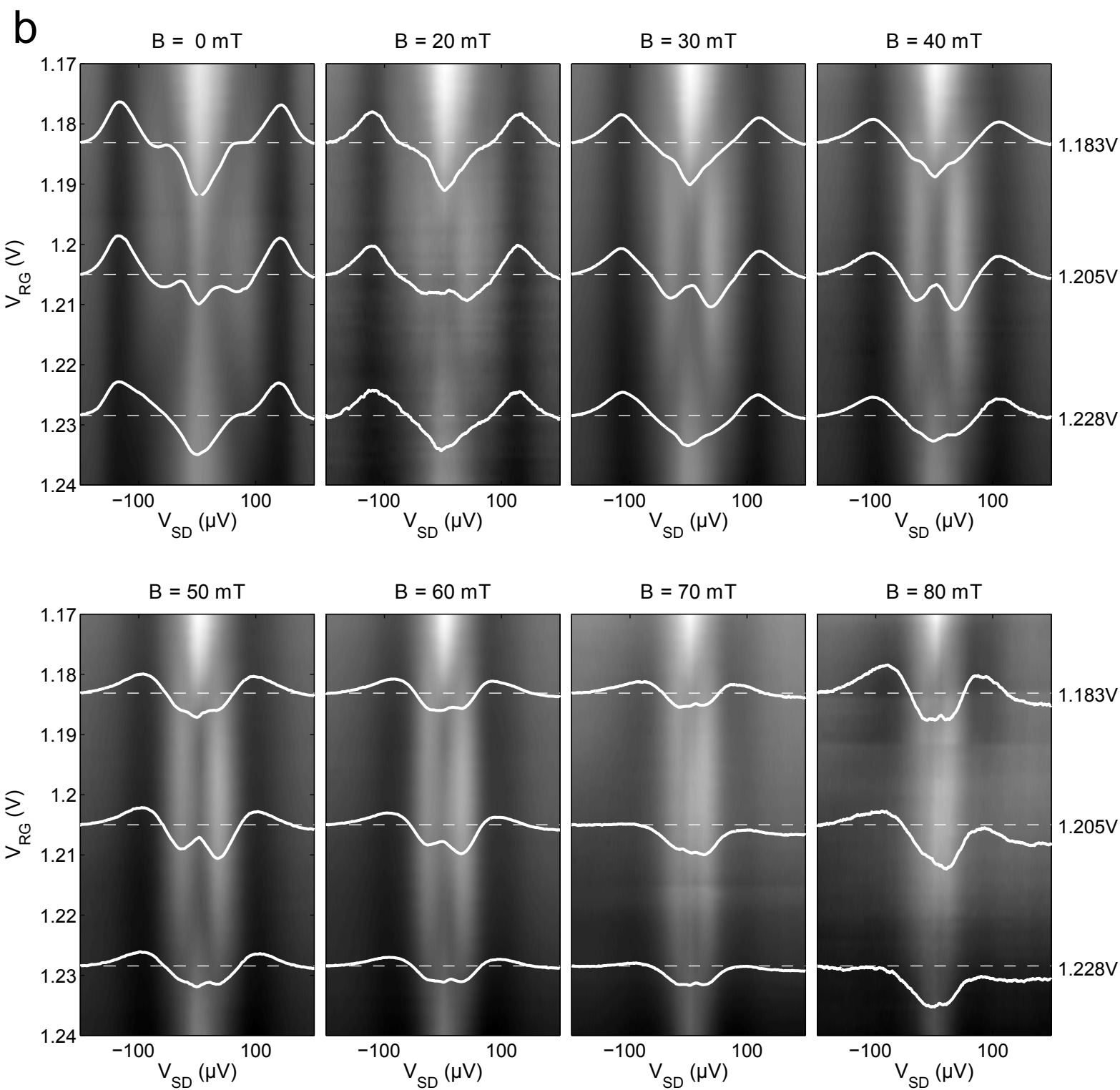

S8

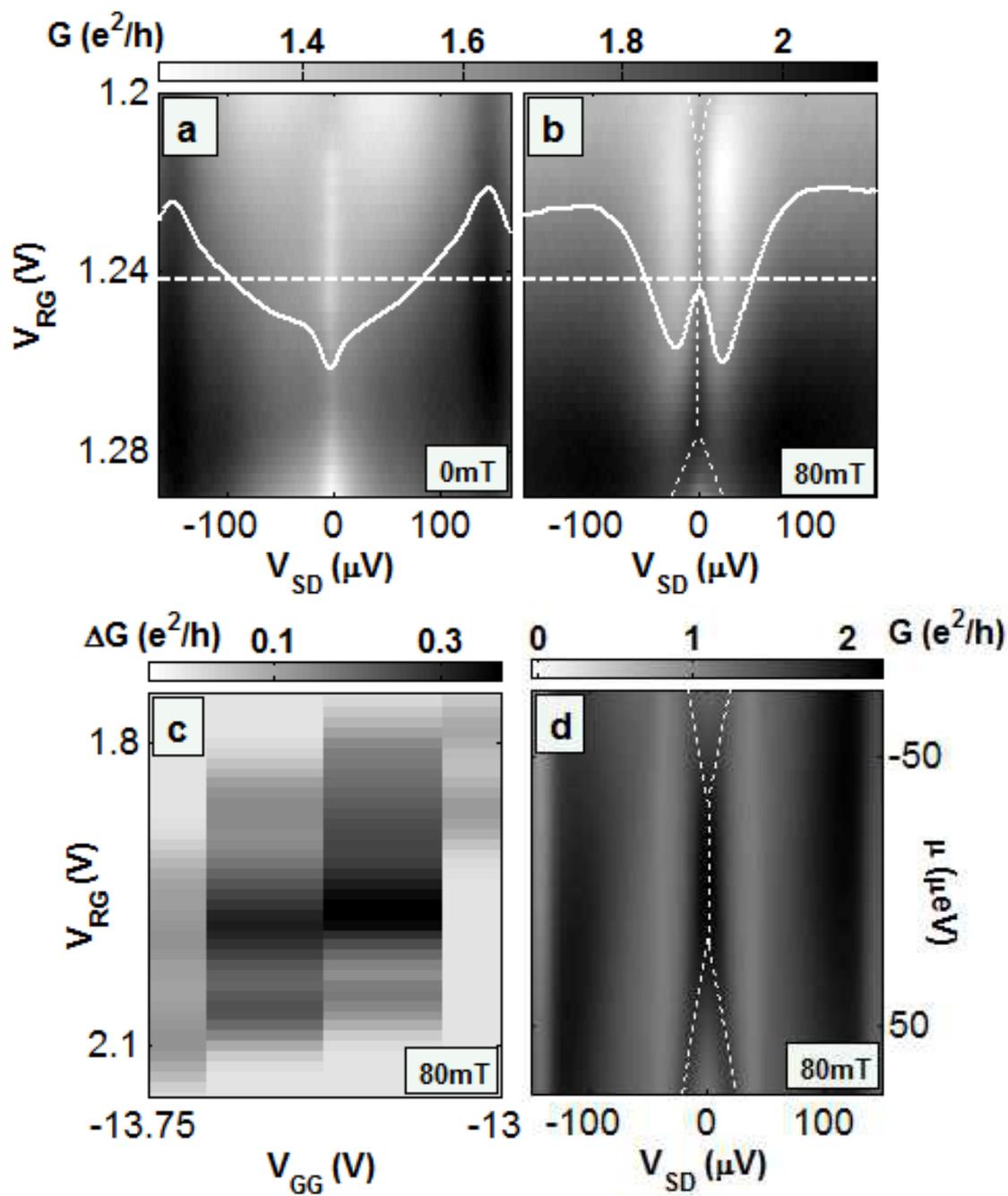

S9

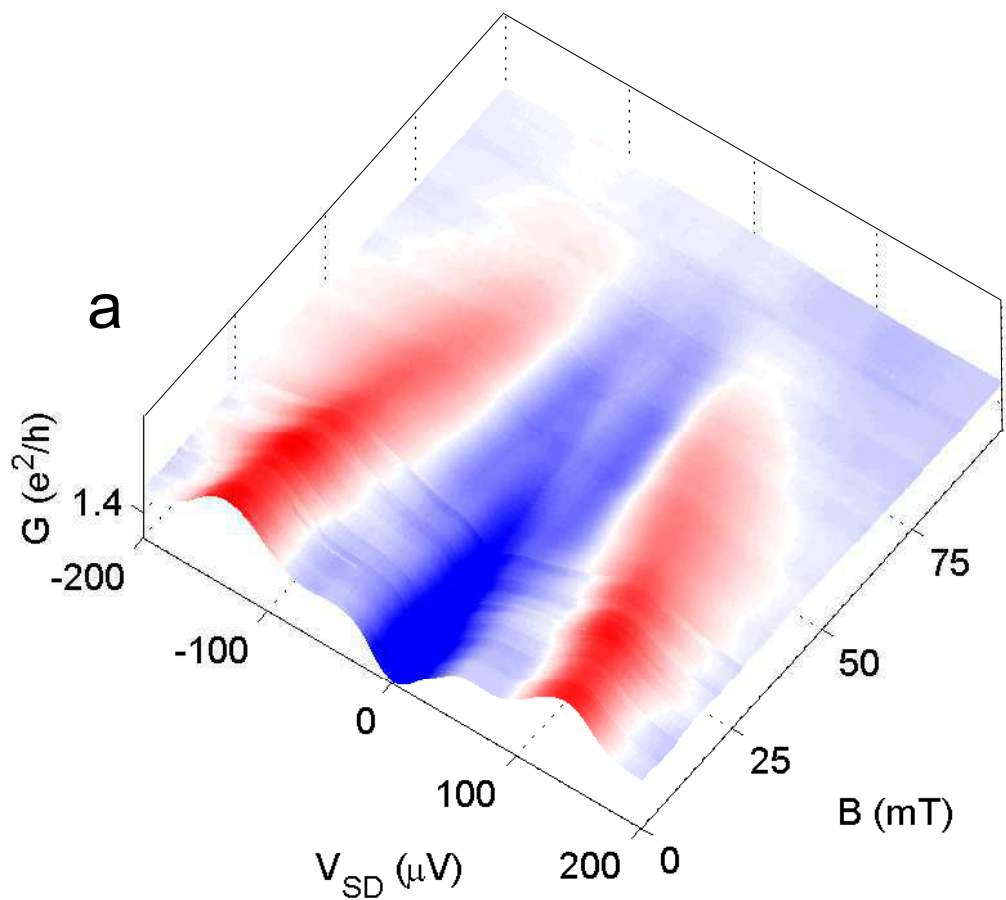

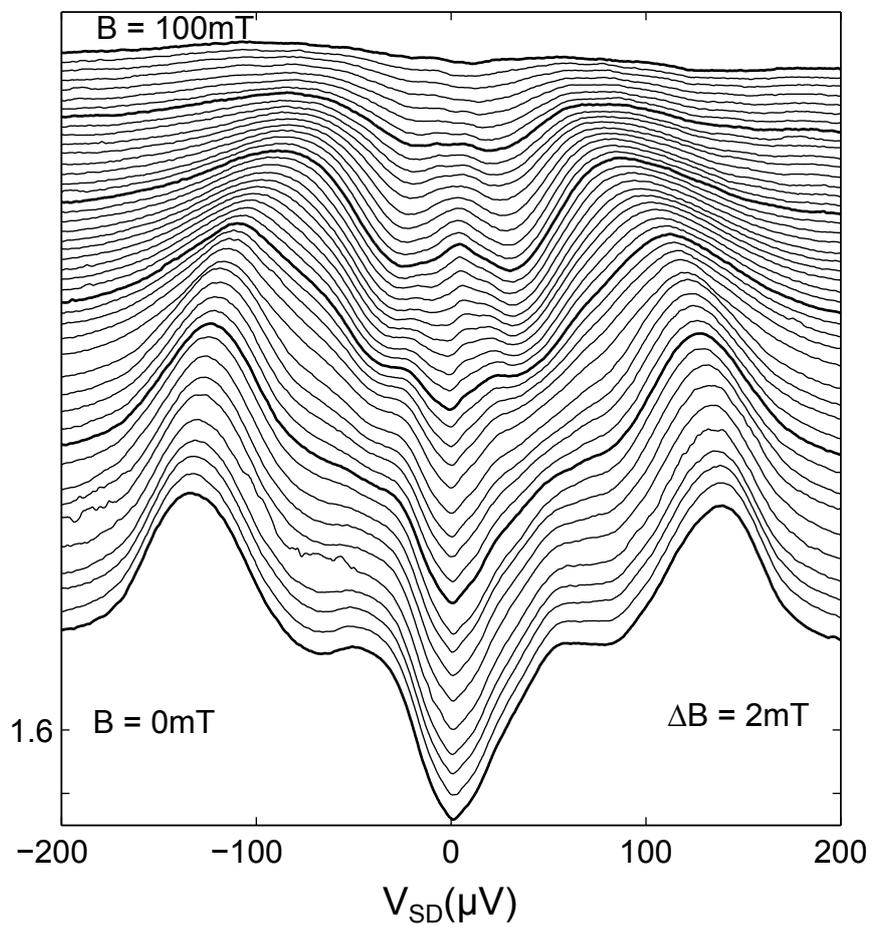

S10

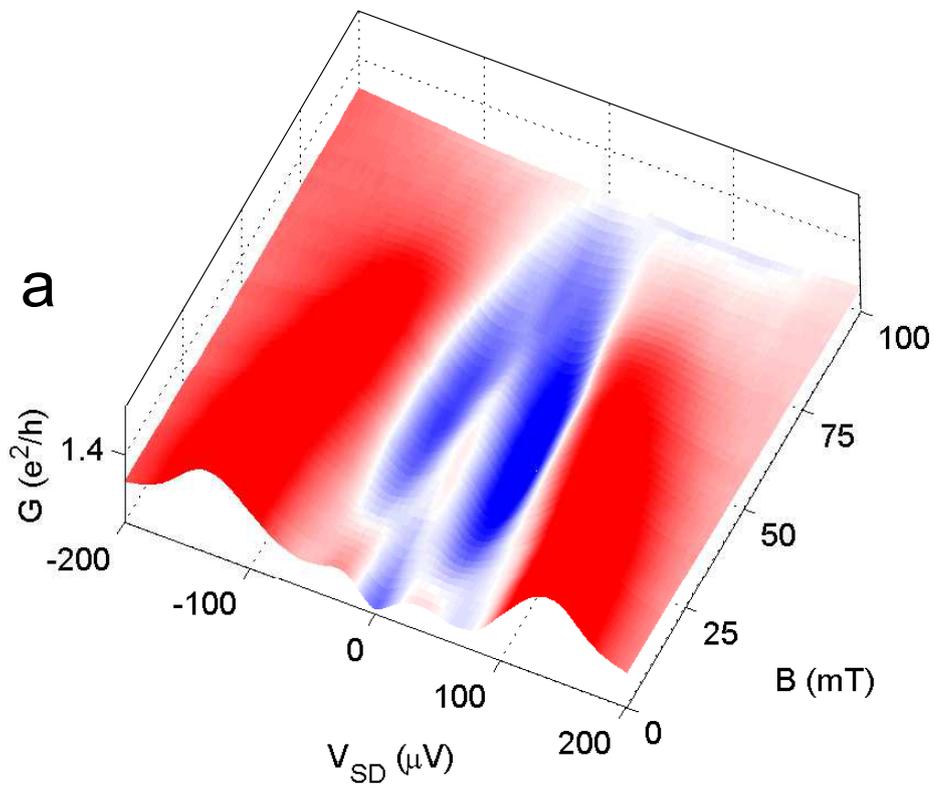
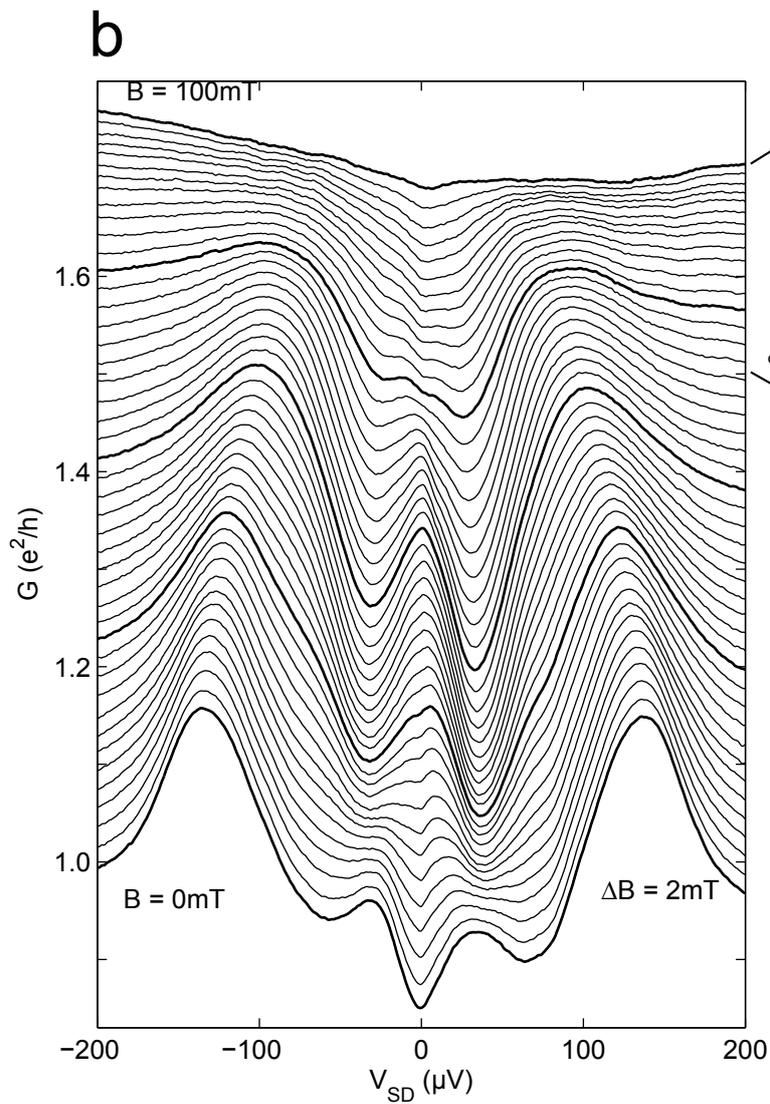
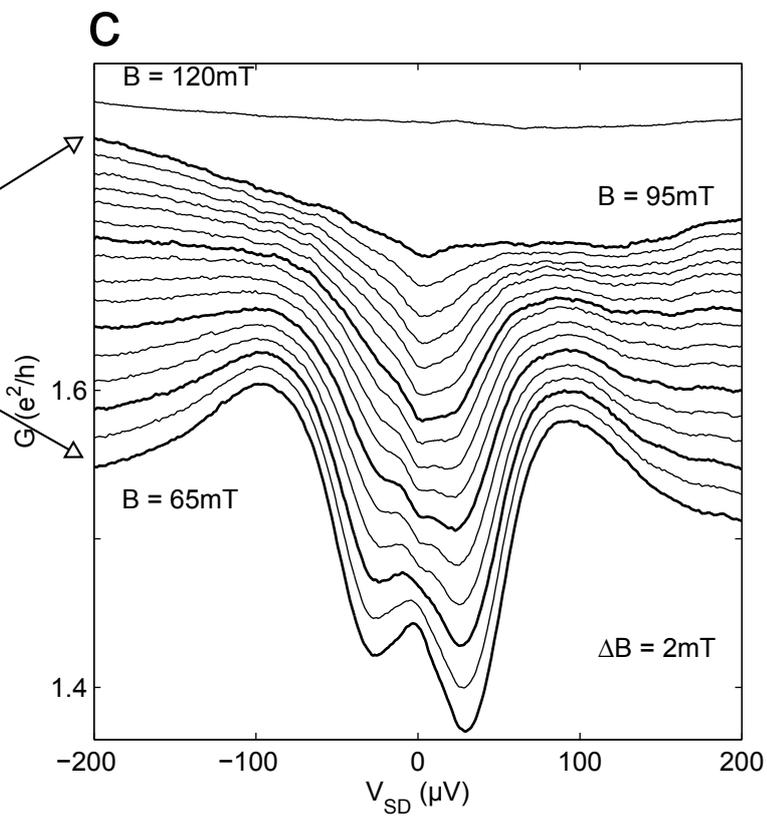

S11

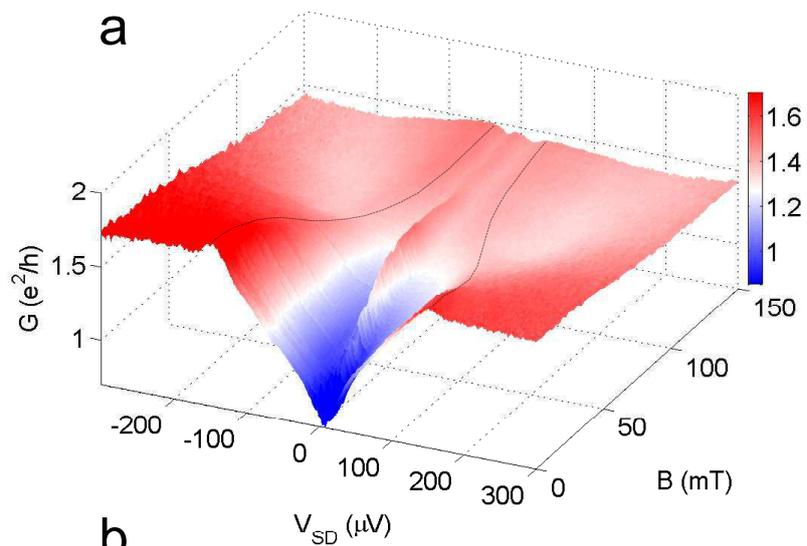
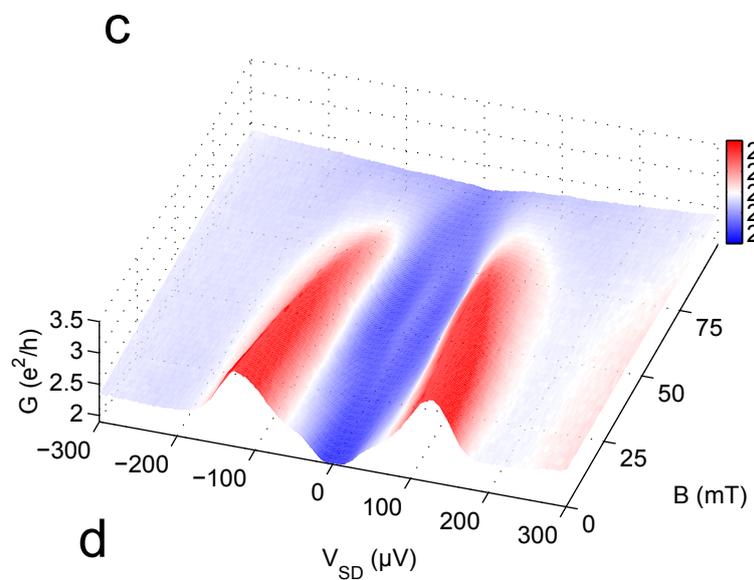
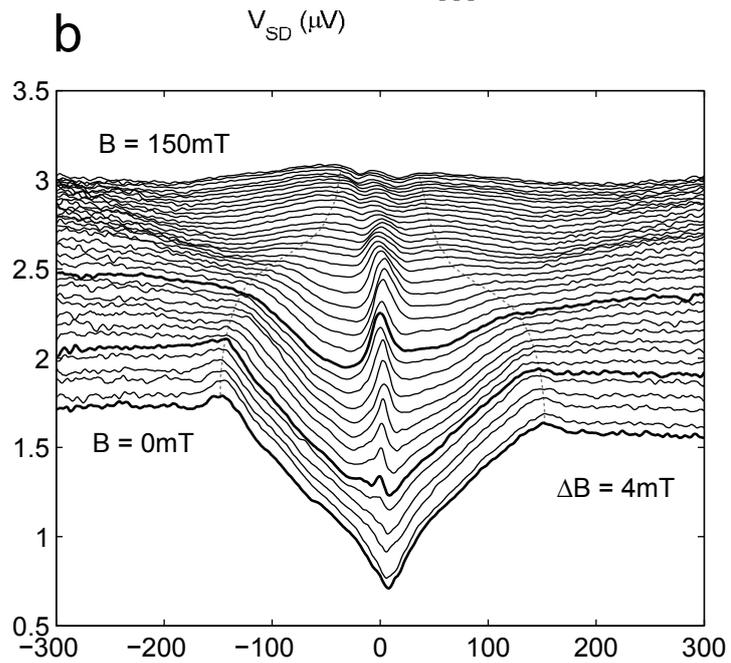
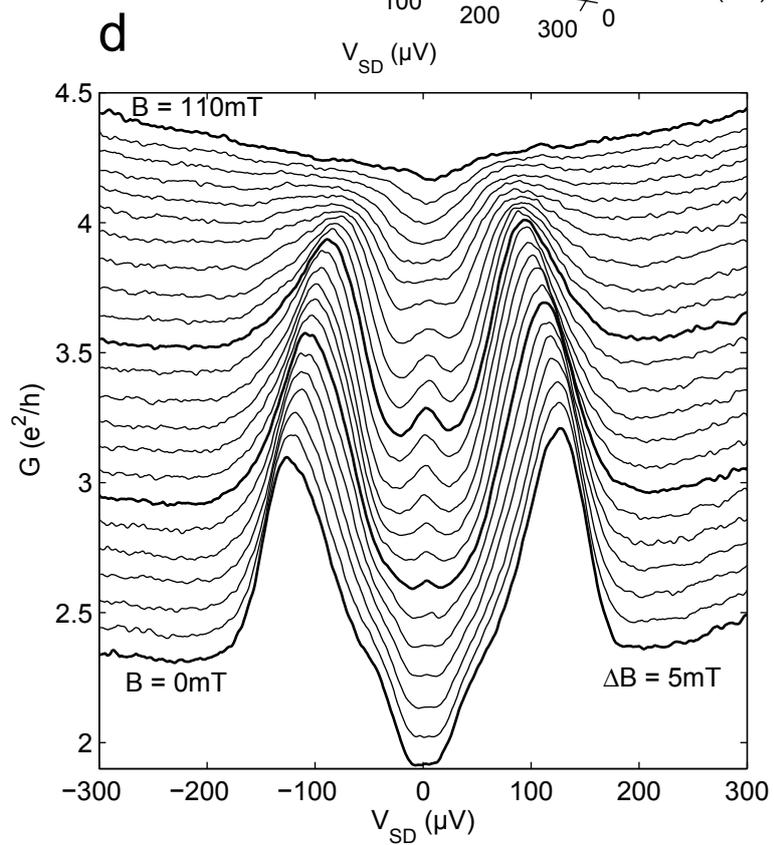